# Water-induced buoyancy controls transient water storage in the mantle transition zone

T. V. Gerya[1], N.M. Bardi[1], S. Karato[2] & M. Murakami[1]

[1]Swiss Federal Institute of Technology Zurich, Department of Earth and Planetary Sciences, Zurich, Switzerland.

[2]Yale University, Department of Earth & Planetary Sciences, New Haven, CT, United States

**The spinel phase (wadsleyite, ringwoodite) in the mantle transition zone (MTZ), can contain up to 1–2 wt% of water[1,2]. However, whether these water reservoirs in the MTZ are filled is debated[2,3]. Here, we investigate water dynamics in the MTZ numerically by using a newly developed empirical model of deep hydrous mantle melting combined with 2D thermo-hydro-mechanical-chemical[4] (THMC) upper mantle models. Numerical modeling results suggest that water-induced buoyancy triggers the development of hydrous solid-state mantle upwellings in the MTZ[5,6]. On time scales of some tens of millions of years, they rise to and interact with the spinel-olivine phase transition. Depending on the water content and temperature of these thermal-chemical plumes, this crossing may trigger hydrous melting by water release from the wadsleyite upon its conversion to olivine[3,7]. The melts are less dense than the solid matrix and continue rising upward in the form of either diapirs or porosity waives. Similar dehydration-induced melting process[3] is also documented for the lower MTZ boundary, where hydrous downwellings (such as subducted slabs) generate buoyant melt diapirs rising through the MTZ. We therefore suggest that the MTZ operates as a transient water reservoir. Relatively small amounts of water (<0.1 wt%, <0.2 ocean masses) and a geologically moderate duration (80-430 Myr) of the transient water storage should be characteristic for the MTZ, which may play a key role in stabilizing the surface ocean mass on Earth and Earth-like rocky exoplanets[2,8].**

The long-term stability of surface ocean on Earth and Earth-like rocky exoplanets is a prerequisite for maintaining habitable conditions and is governed by interactions with the mantle[3]. The water capacity of mantles for rocky planets is dominated by minerals in the mantle transition zone (MTZ). These spinel phase minerals (wadsleyite, ringwoodite) can contain up to 1–2 wt% of water[1,2] well beyond the capacity of other minerals. However, whether these water reservoirs in the MTZ of rocky planets are filled is debated[2,3]. For Earth, water content estimates in the MTZ range from less than 1 to up to 11 ocean masses[1], which can be strongly dependent on the water dynamics in the MTZ and, in particular, the efficiency of the transition-zone water filter effect[7]. This effect implies that the hydrated mantle ascending through the top boundary of the MTZ undergoes dehydration-induced partial melting[7,9]. Depending on the mantle composition, the formed Fe-rich hydrous melt may potentially be denser than the surrounding solid and accordingly trapped at the top MTZ boundary, which may thus delay or even preclude the efficient water return to the surface[7]. It has also been hypothesized that the filter effect could be suppressed for mantle plumes and a hotter mantle potential temperature[7].

An important additional unstudied issue is the nature and dynamics of the similar water filter effect associated with down-going mantle flow at the lower MTZ boundary, related for example to downward penetrating hydrated slabs through this boundary. Similar to the processes at the MTZ-upper mantle boundary, if water-rich MTZ materials move into the lower mantle, then hydrous partial melting may occur, and if the melts were less dense than the surrounding solid[3], hydrous melts will leak back into the overlaying reservoir. And thus, the geophysically inferred water-rich MTZ state[10] can be dynamically stimulated. It is thus important to consider the processes in both upper and the lower MTZ boundaries to understand the nature of global water circulation[3,7].

The processes of when and how water and hydrous melt will migrate below, inside and above the MTZ remain rather poorly understood. In turn, water dynamics in the MTZ of rocky (exo)planets with surface oceans critically affects its ability of these planets to maintain the long-term stability of their surface ocean volume. In the presence of continents (i.e., significant subaerial land masses), the surface ocean level stability is likely a pre-requisite for maintaining stimulating habitable conditions for complex and advanced life[11], which needs careful understanding and quantification. Even for Earth, the long-term stability of the surface ocean volume and emergence of continents above the sea level remain controversial[12,13] and more uncertainty exists for other terrestrial planets and rocky exoplanets.

Previous numerical studies[5,6] predicted that thermal relaxation of cold, water-rich downwellings (such as subducting slabs and lithospheric drips) accumulating in the bottom of the MTZ should trigger the development of cold hydrous upwellings (plumes, diapirs) moving upward from this region, which are driven by the water-induced buoyancy[5,6]. These models however did not explore the effects of hydrous mantle melting, which leave important uncertainties concerning the extent and efficiency of MTZ water filter effects both above and below this region[3,7].

In this paper, we investigate thermodynamically and numerically the water dynamics in the MTZ by developing a new experimentally-based empirical model of deep hydrous mantle melting combined with 2D numerical thermo-hydro-mechanical-chemical (THMC) MTZ models with solid-state phase transitions and hydrous melt formation and percolation in the mantle. Our results indicate that water-induced buoyancy of hydrous mantle melts will likely overcome effects of Fe-incorporation into melts. MTZ water storage for Earth-like planets will therefore be transient. This should define strong MTZ undersaturation in term of water content as well as promote the stability of surface ocean volume, which cannot be efficiently recycled into the MTZ by subduction processes.

**Deep hydrous mantle melting**

In order to constrain water dynamics in the MTZ region by THMC models, thermodynamic calculations are required[4], which couple composition, density and volumetric proportions of coexisting melt and solid matrix thereby taking into account both volumetric and thermal effects of melting and solid-state phase transformations. Such a thermodynamic framework is currently unavailable mainly because the existing experimentally-based mantle melting models[14,15] are calibrated with relatively low-pressure experiments and do not incorporate available experimental data for melting at MTZ P-T conditions[16-20]. To overcome this deficiency, we calibrated a new experimentally-based empirical model of deep hydrous mantle melting (see Methods), which can account in a simplified manner for compositional variations of coexisting hydrous melt and solid matrix under characteristic P-T conditions of the MTZ. The melting model is coupled to calculations of coexisting solid matrix and hydrous melt density on the basis of a thermodynamic database for the deep mantle[21], a calibrated reaction of water incorporation from coexisting melt into wadsleyite (Methods), known water partitioning coefficients from wadsleyite to ringwoodite, olivine and perovskite[22] and the equations of state for the hydrous melt[23] and water[24].

One important question related to water stability in the MTZ is related to variations of density difference between the coexisting solid matrix and hydrous melt. In particular, the transition-zone water filter effect has been proposed[3,7] by analyzing consequences of very different water capacities of nominally anhydrous mantle minerals inside and outside MTZ.

Fig. 1 shows results of thermodynamic modeling of the density difference between coexisting solid and melt for pyrolite mantle at the MTZ P-T conditions. As follows from this figure, melt density remains below the density of the solid matrix in the entire MTZ P-T range irrespective of the water content in the mantle. Below the lower MTZ boundary, this difference can reach up to ca. 600 kg/m$^3$, which suggest efficiency of upward return flow of produced hydrous melts back into the MTZ[3]. Above the upper MTZ boundary, solid-melt density difference reduces drastically (up to ca. 200 kg/m$^3$) mainly because of the wadsleyite-olivine phase transition in the solid matrix[7]. The hydrous melt, however, remains positively buoyant relative to the solid matrix and should therefore move upward rather than downward, thereby controlling gradual water loss from the MTZ through its upper boundary. Melt-solid density contrast depends notably on the iron content in the mantle and may further reduce for more iron-rich compositions (Extended Data Figure 3). This compositional effect should control water dynamics in the interior of rocky exoplanets with variable mantle compositions[2,8].

The new empirical model of hydrous mantle melting is coupled to 2D THMC upper mantle model (see Methods), which allowed us to simulate water dynamics in the entire range of P-T conditions below, inside and above the MTZ. We performed simulations in 5 different geodynamic situations relevant for the MTZ water cycle[3,5-7]: (1) upward hydrous plume penetration through the upper MTZ boundary across the wadsleyite-olivine phase transition, (2) development of gravitational instability at the upper MTZ boundary caused by uniformly hydrated MTZ, (3) downward hydrated cold dense drip penetration through the lower MTZ boundary across the ringwoodite-perovskite phase transition, (4) upward penetration of hot hydrous plume through the lower MTZ boundary and (5) large-scale dynamics of cold hydrated slab fragments located in the bottom of the MTZ. Results from these experiments consistently predict transient nature of water storage in the MTZ controlled by the water-induced mantle buoyancy and related thermal-chemical convection and upward hydrous melt percolation and assembling.

### Hydrated plume at the upper MTZ boundary

Fig. 2 show dynamics of cold hydrous upwelling approaching the MTZ upper boundary, which can be formed by thermal relaxation and related rheological softening of stagnated subducted slabs[5,6]. This thermal-chemical plume with initial cold temperature anomaly of -100 K is driven upward by the water-induced chemical buoyancy[5,25] (Fig. 2a,e,i). A positive Clapeyron slope characteristic of the spinel-olivine transition causes the cold upwelling to slow down and flatten at the MTZ boundary until its temperature rises enough to cross the transition (Fig. 2b,c). This happens as a positive Clayperon slope implies that lower pressure (and depth) is required to stabilize wadsleyite inside the colder plume compared to the surrounding warmer ambient MTZ mantle, a previously unrecognized geodynamic effect. The reduction of plume positive buoyancy created by this effect slows down or even temporarily arrest the upward plume motion. This thermodynamic effect is to some degree analogous to the well-known mechanism causing cold slab stagnation at the lower MTZ boundary. There, a negative Clapeyron slope of the spinel-perovskite transition requires higher pressure (and depth) of the transition inside the descending cold slab compared to the warmer ambient mantle.

At sufficiently high water content in the plume ($\geq$0.3 wt%), the spinel-olivine phase transition causes dehydration-induced melting due to the lower water capacity of olivine compared to wadsleyite (Fig. 1). At Earth-like mantle composition, the density of resulting Fe-rich hydrous melt is by 150-200 kg/m$^3$ lower than the density of the coexisting solid (Fig. 1). This density contrast enables melt rising toward the surface in form of either porosity waves or in form of a diapir. The choice of melt upwelling mode is mainly dependent on the water (and melt) content in the ambient mantle located above the MTZ: higher (>0.3%) water content stabilizes porous melt in the mantle and enables porosity waves (Extended Data Fig. 3) whereas lower (<0.3 wt%) water content precludes hydrous melting of the ambient mantle and promotes rising of a complex plume/diapir with melt being concentrated on its leading edge (Fig. 2g,h).

**Gravitational instability of the uniformly hydrated MTZ**

Extended Data Fig. 4 shows the density profile of uniformly hydrated MTZ, which apparently look gravitationally stable as the density of hydrated spinel-bearing mantle remans higher than the density of overlaying drier olivine-bearing mantle. This density profile is perturbed by a small circular temperature anomaly of +10 K at the upper MTZ boundary. On a long term, this perturbation leads to the development of exponentially growing gravitational instability across the phase transition, which produces a large-scale overturn of the hydrated MTZ and drier overlaying mantle. The Rayleigh-Taylor instability develops exponentially from the small initial thermal perturbation of the density field leading to the development of a hydrated diapir. At high MTZ water content (0.4 wt%), the diapir is subjected to hydrous partial melting upon crossing spinel-olivine transition. Partial melt accumulates in the leading edge of the diapir, creating a thin melt-rich layer. The timescale of development of the instability increases with decreasing initial MTZ water content from ca. 9 Myr for 0.4 wt% (Extended Dara Fig. 3b) to 37 Myr for 0.1 wt% (model diapir91, Extended Data Table 2). This implies that only very low water content can remain in the MTZ for protracted time of some tens of million years.

**Hydrated drip penetration through the lower MTZ boundary**

Fig. 3 shows initial conditions and results for the experiment simulating consequence of cold hydrated drip penetration through the lower MTZ boundary defined by the spinel-perovskite transition. The driving force for this penetration resides in the higher density of the fragment core due to the presence of an Fe-rich circular drip core mimicking fertilized mantle. This situation is somewhat different from that of hydrated subducting slab penetration[3] driven by the thermal (rather than compositional) negative slab buoyancy in the upper mantle. However, penetration-induced melting processes will remain largely similar, which approve the use of this highly simplified model setup. Upon penetration, hydrous melts produced inside the drip percolate upward toward its rear part. As the result, a high-degree localized lens-like melting region (>10 vol.% melt) forms inside (Fig. 3f) the downward moving drip and stays behind it (Fig. 3g). This positively buoyant melt continues to focus and penetrates back into the MTZ in form of two small diapirs (<3 km in diameter), where it continues rising upward toward the upper MTZ boundary without being solidified (Fig. 3h). As such, hydrous melt produced by the slab penetration does not remain at the lower MTZ boundary but is rapidly transported across the MTZ. The downward drip penetration is mainly driven by the compositional negative buoyancy. Drips less enriched in iron and with smaller enriched core size (cf. models diapir69, diapir71, diapir73, diapir78, Extended Data Table 2) tend to stagnate above the spinel-perovskite transition with only partial or no penetration into it.

**Hydrated plume at the lower MTZ boundary**

Fig. 4 shows hot hydrous plume interaction with the lower MTZ boundary. Positive temperature anomaly of the plume and negative Clapeyron slope of the spinel perovskite phase transition drive plume stagnation at this boundary. Due to the low water capacity of perovskite, hydrous melt instantaneously form inside the plume and start to move upward by porosity waves (Fig.4f) forming a thin melt-rich layer (>3 vol% melt) at the spinel-perovskite transition. This layer destabilizes by producing small (<2 km in diameter) melt diapir (>90 vol% melt) rapidly rising across the MTZ (Fig.4 g,h). The depleted compositional core of the plume stagnates and flattens below and within the spinel-perovskite transition. The plume's core stagnation depends on the size of the core. A larger depleted core size leads to plume penetration into and ascent within the MTZ (Extended Data Fig. 5). In this case, the partially molten plume below the MTZ with porosity waves inside evolves into a solid plume inside the MTZ with melt lens on top. In contrast, a non-depleted, hot hydrous plume rapidly focuses melt in its top part so that a small secondary melt diapir (2-3 km in diameter, >90 vol.% melt) is formed, which rises more rapidly than the plume itself. Whereas the hot plume is arrested by the negative Clapeyron slope of perovskite-ringwoodite phase transition, the melt diapir continues to rise through the MTZ unimpeded (cf. model diapir89 in Extended Data Table 2). In this way, hydrous melt is efficiently transported upward and can reach the upper MTZ boundary, subsequently rising further toward the surface.

**Large-scale dynamics of cold hydrated slab fragments**

Extended Data Fig. 6 shows dynamics of an elongated cold hydrated slab fragment located above the lower MTZ boundary. Initially, due to the cold temperature, the density of the fragment is larger than that of the ambient warmer and drier MTZ mantle. Therefore, the fragment moves downward until colliding with the lower MTZ boundary, where it stagnates and relaxes thermally for ca. 30 Myr (Extended Data Fig.6d) until its negative temperature anomaly reduces to less than -100 K. In agreement with previous numerical studies[5,6], thermal relaxation of the slab fragment triggers solid-state hydrous plumes from its warmer edges, which are driven by the water-induced buoyancy. In addition, partial penetration of the hydrated fragment through the spinel-perovskite transition triggers dehydration-induced melting which form small melt diapirs on two edges of the deformed slab material (Extended Data Fig.6d). Solid state hydrous plumes and melt diapirs rise together across the MTZ on timescales of some tens Myr fragment (Extended Data Fig.6g). The positive Clapeyron slope of the ringwoodite-wadsleyite transition, which is located in the middle of the MTZ, causes transient delay of the cold plumes rise, similarly to what has been already described for the wadsleyite-olivine transition. Arrival of the cold hydrous plumes to the upper MTZ boundary results in yet further delay and triggers hydrous melting similarly to the first series of experiments (cf. Extended Data Fig. 6j and Fig.2c). A twin model with a smaller initial negative temperature anomaly of -100 K (Extended Data Fig. 7a) does not produce significant slab fragment subsidence into the spinel-perovskite transition region. As a result, a larger single heterogeneous solid state plume develops, which moves more rapidly across the MTZ within 60-70 Myr (Extended Data Fig. 7d,g). Penetration of this plume across the spinel-olivine transition is analogous to previous models and triggers dehydration-induced melting resulting in the development of melt lenses on the leading edge of the plume (Extended Data Fig. 7j-l). Similar single solid state plume also forms in the case of a large depleted slab core (cf. models diapir83, diapir84 in Extended Data Table 2), which moves more rapidly across the MTZ within ca. 20-30 Myr.

The timescales of large-scale water dynamics in the MTZ are thus mainly defined by the slab relaxation time at the lower MTZ boundary (ca. 20-30 Myr in our models) and hydrous plumes and diapirs rising time across the MTZ (ca. 20-90 Myr in our models), which in turn depend on their characteristic water- and depletion-induced chemical buoyancy and thickness. As the result, these timescales can vary from 50 to >150 Myr, thereby defining the magnitude of the transient water residence time in the MTZ.

**Density crossover and MTZ water filter effect**

Density crossover corresponds to conditions at which density of the mantle-derived melt becomes larger than the density of coexisting solid phase and this effect has originally been proposed to explain differentiation of the terrestrial magma ocean[26-28]. It has been demonstrated experimentally that the density crossover of the dry peridotite melt and the equilibrium olivine occurs in the range of pressure and temperature immediately above the MTZ, where olivine should float in the coexisting primitive peridotite melt[26-29]. Later, density crossover has been also hypothesized for hydrous mantle melts forming above the MTZ as the result of dehydration melting[3,7,30]. Some experiments on hydrous mantle melting[31] indeed suggest strong enrichment of low-degree (2-7 vol%) hydrous melts in iron, which can notably increase the melt density. However, some other recent experiments show rather moderate iron enrichment of low-degree (7 vol%) hydrous melt formed at low water content in the peridotite[20].

      Bercovici and Karato[7] proposed that when the ascending ambient hydrated mantle rises out of MTZ into the low-solubility upper mantle above 410 km it undergoes dehydration-induced partial melting that filters out incompatible elements. The filtered, dry and depleted solid phase continues to rise whereas the wet, enriched hydrous melt may be denser than the surrounding solid matrix and accordingly trapped at the MTZ upper boundary[7]. The potential feasibility of the density crossover has been tested experimentally by varying water content in the peridotite melt under MTZ P-T conditions and showing that at certain critically low water content in the melt its density can approach that of the solid peridotite[23,30,31]. Our experimentally based empirical melting model however suggests that, for hydrous pyrolite mantle compositions, the equilibrium water content in the hydrous melt in the MTZ temperature range of 1400-2000°C always remains higher than the critical water content and the density crossover does not occur. Density of the hydrous partial melt remains lower than the density of coexisting solid matrix (composed of olivine, pyroxenes and majoritic garnet) by at least 150-200 kg/m$^3$ (Fig. 1). The density contrast decreases with increasing concentration of iron-rich component B (Extended Data Table 1) in the mantle (cf. Fig. 1 and Extended Data Fig. 1) and the density crossover is predicted to occur at $X_B$=0.67, which is four times more enriched mantle composition compared to pyrolite. This prediction has however to be taken with caution since our melting model is calibrated for pyrolite bulk mantle composition and may not be accurate for very different mantle compositions.

In order to test potential dynamical effects of both neutrally (i.e., at the density crossover) and strongly negatively buoyant hydrous melts, we conducted four numerical experiments similar to the model shown in Fig.2 but with an ad hock increment added to the computed melt density of 150, 200 kg/m$^3$ (neutrally buoyant melt) and 300, 400 kg/m$^3$ (negatively buoyant melt), respectively. These experiments suggest that the neutrally buoyant melt created by dehydration melting above the MTZ is passively transported upward by chemically buoyant hydrous mantle plumes driven by the water-induced density reduction. In contrast, strongly negatively buoyant melt tends to stand behind the plume and then slowly return to the top of the MTZ where it can reside (Extended Data Fig. 8), thereby retaining some part of the initial water present in the rising plume, similarly to what has been hypothesized by Bercovici and Karato[7]. The feasibility of this effect has however to be tested more rigorously for variable mantle compositions and water contents. This in turn requires high-accuracy experimental data for equilibrium low-degree hydrous partial melt compositions[20,31] as well as calibration of new experimentally based MTZ melting model valid for non-pyrolite mantles.

**Dynamically supported water content in the MTZ**

Based on the results of our solid/melt density calculations and numerical experiments, we suggest that, due to the intrinsic positive buoyancy of (i) hydrated solid mantle compared to dry mantle and (ii) hydrous melts compared to coexisting solid matrix, the MTZ operates as a transient water reservoir (Fig. 5). Water content in the MTZ should therefore be regulated by the balance between incoming and outgoing water on characteristic water retention timescale. This timescale is mainly dictated by (i) the characteristic time of thermal relaxation of hydrated slabs and drips ($t_{relaxation}$=70-330 Myr, depending on the slab thermal thickness) and (ii) the characteristic time needed for hydrous upwellings to leave the MTZ ($t_{escape}$=10-100 Myr, depending on the water content and size, based on our models). For modern Earth, the global balance of water on these timescales can be represented as

$$WATER_{MTZ} = water_{subduction}(t_{relaxation} + t_{escape}),$$

$$t_{relaxation} = \frac{h^2}{\kappa},$$

where $water_{subduction}$= 2–6 × $10^{17}$ kg/Myr[33-35] is water mass subducted to MTZ per million year, $h$ = 45-100 km is thermal thickness of subducted slabs[32], $\kappa$ = 30 km$^2$/Myr ($10^{-6}$ m$^2$/s) is thermal diffusivity. Relatively small amount of water (0.004-0.06 wt%, 0.01-0.19 ocean masses), consistent with geophysical estimates[10], and geologically moderate duration (80-430 Myr) of the transient water storage should be characteristic for the MTZ, which may thus play a key role in stabilizing surface ocean level on Earth and Earth-like rocky exoplanets. This relatively low average global MTZ water storage estimate is in strong contrast with estimated large values based on the water capacity of mantle minerals[2], which in particular suggest a key role of MTZ water dynamics for maintaining the long-term planetary habitability.

Our models also suggest that water dynamics above the MTZ can alternate between porosity waves and melt diapirs, with the mode depending critically on the ambient water content and olivine water capacity. As the result, the hydrous melt transport efficiency, and thus the overall escape time of water from the MTZ, is highly sensitive to a switch in transport mode governed by the upper mantle's water content. To solidify these estimates, future work should rigorously address the sensitivity of the mode-switching conditions to fundamental physical parameters. Specifically, conducting further sensitivity analyses on how melt viscosity and the solid matrix's permeability and rheology influence the transition from porous flow to diapir formation will be an important step in refining the constraints on water transport efficiency and residence time within the MTZ (Fig. 5), which in particular impose geodynamical and geochemical controls on intraplate volcanism[36-38].

**References**

1. Ohtani, E. Hydration and Dehydration in Earth's Interior. Annual Rev. Earth Planet. Sci. 49, 253-278 (2021).
2. Guimond, C.M., Shorttle, O., Rudge, J.F. Mantle mineralogy limits to rocky planet water inventories. Monthly Notices Royal Astronom. Soc. 521, 2535-2552 (2022).

3. Karato, S., Karki, B., Park, J. Deep mantle melting, global water circulation and its implications for the stability of the ocean mass. Progress Earth Planet. Sci. 7, 76 (2020).
4. Gerya T.V. Introduction to Numerical Geodynamic Modelling. Second Edition. Cambridge University Press, 472 pp (2019).
5. Richard, G. C., Bercovici, D. Water-induced convection in the Earth's mantle transition zone. J. Geophys. Res. 114, B01205 (2009).
6. Perchuk, A. L., Gerya, T. V., Zakharov, V. S. Griffin, W. L. Building cratonic keels in Precambrian plate tectonics. Nature, 586, 395-401 (2020).
7. Bercovici, D., Karato, S. Whole-mantle convection and the transition zone water filter. Nature 425, 39–44 (2003).
8. Spaargaren, R. J., Wang, H. S., Mojzsis, S. J., Ballmer, M. D., Tackley, P.J. Plausible constraints on the range of bulk terrestrial exoplanet compositions in the Solar neighborhood. Astrophys. J. 948, 53 (2023).
9. Ohtani, E., Litasov, K., Hosoya, T., Kubo, T., Kondo, T. Water transport into the deep mantle and formation of a hydrous transition zone. Phys. Earth Planet. Interiors 143–144, 255–269 (2004).
10. Karato, S. Water distribution across the mantle transition zone and its implications for global material circulation. Earth Planet. Sci. Let. 301, 413–423 (2011).
11. Stern, R. J., Gerya, T. V. The importance of continents, oceans and plate tectonics for the evolution of complex life: implications for finding extraterrestrial civilizations. Sci. Rep. 14, 8552 (2024).
12. Korenaga, J. Was there land on the early Earth? Life 11, 1142 (2021).
13. Miyazaki, Y., Korenaga, Y. A wet heterogeneous mantle creates a habitable world in the Hadean. Nature 603, 86–90 (2022).
14. Katz, R. F., Spiegelman, M., Langmuir, C. H. A new parameterization of hydrous mantle melting. Geochem. Geophys. Geosyst. 4, 1073 (2003).
15. Holland, T.B., Green, E.C.R., Powell, R. Melting of peridotites through to granites: A simple thermodynamic model in the system KNCFMASHTOCr. J. Petrology 59, 881–900, (2018).

16. Inoue, T., Sawamoto, H. High pressure melting of pyrolite under hydrous condition and its geophysical implications. In: *High-Pressure Research. Application to Earth and Planetary Sciences* (ed. Y. Syono, M. H. Manghnani) 323 – 331 (Geophysical Monograph Serie, AGU, Washington D.C., 1992).
17. Zhang, J., Herzberg, C. Melting experiments on anhydrous peridotite KLB-1 from 5.0 to 22.5 GPa. J. Geophys. Res. 99, 17,729-17,742 (1994).
18. Litasov,K., Ohtani, E. Phase relations and melt compositions in CMAS–pyrolite–H2O system up to 25 GPa. Phys. Earth Planet. Interiors 134, 105–127 (2002).
19. Mibe, K., Orihashi, Y., Nakai, S., Fujii, T. Element partitioning between transition-zone minerals and ultramafic melt under hydrous conditions. Geophysical Research Letters, 33, L16307 (2006).
20. Fei, H., Chen, J., Wang, F., Zhang, B., Xia, Q., Katsura, T. Gravitationally unstable hydrous melts at the base of the upper mantle. J. Geophys. Res. 130, e2024JB030737 (2025).
21. Stixrude, L., Lithgow-Bertelloni, C. Thermal expansivity, heat capacity and bulk modulus of the mantle. Geophys. J. Int. 228, 1119–1149, (2022).
22. Inoue, T., Wada, T., Sasakia, R., Yurimoto, H. Water partitioning in the Earth's mantle. Phys. Earth Planet. Interiors 183, 245–251 (2010).
23. Jing, Z., Karato, S. Effect of $H_2O$ on the density of silicate melts at high pressures: Static experiments and the application of a modified hard-sphere model of equation of state. Geochim. Cosmochim. Acta 85, 357–372 (2012).
24. Gerya T. V., Podlesskii K. K., Perchuk L. L., Maresch, W. V. Semi-empirical Gibbs free energy formulations for minerals and fluids. Phys. Chem. Minerals 31, 429-455 (2004).
25. Gerya, T. V., Yuen, D. A. Rayleigh-Taylor instabilities from hydration and melting propel "cold plumes" at subduction zones. Earth Planet Sci. Let. 212, 47-62 (2003).
26. Agee. C. B., and Walker, D. Static compression and olivine flotation in ultrabasic silicate liquid. J. Geophys. Res. 93, 3437-3449 (1988).
27. Ohtani, E., Nagata, Y., Suzuki, A., Katoa, T. Melting relations of peridotite and the density crossover in planetary mantles. Chem. Geol. 120, 207-221 (1995).

28. Suzuki, A., Ohtani, E. Density of peridotite melts at high pressure. Phys. Chem. Minerals 30, 449–456 (2003).
29. Ohtani, E.. Suzuki, A. and Kato, T. Flotation of olivine in the peridotite melt at high pressure. Proc. Jpn. Acad., Ser. B 69, 23-28 (1993).
30. Matsukage, K. N., Jing, Z., Karato, S. Density of hydrous silicate melt at the conditions of Earth's deep upper mantle. Nature, 438(7067), 488–491 (2005).
31. Freitas, D., Manthilake, G., Schiavi1, F., Chantel, J., Bolfan-Casanova, N., Bouhifd, M.A., Andrault, D. Experimental evidence supporting a global melt layer at the base of the Earth's upper mantle. Nature Comm. 8, 2186 (2017)
32. Garel, F., Goes, S., Davies, D. R., Davies, J. H., Kramer, S. C., Wilson, C. R. Interaction of subducted slabs with the mantle transition-zone: A regime diagram from 2-D thermo-mechanical models with a mobile trench and an overriding plate. Geochem. Geophys. Geosyst., 15, 1739–1765 (2014).
33. van Keken, P. E., Hacker, B. R., Syracuse, E. M., Abers G. A. Subduction factory: 4. Depth-dependent flux of $H_2O$ from subducting slabs worldwide. J. Geophys. Res. Solid Earth https://doi.org/10.1029/2010JB007922 (2011).
34. Faccenda, M., Gerya, T. V., Mancktelow, N. S., Moresi, L. Fluid flow during slab unbending and dehydration: Implications for intermediate-depth seismicity, slab weakening and deep water recycling. Geochem. Geophys. Geosystems 13, Q01010 (2012).
35. Cerpa, N. G., Arcay, D., Padrón-Navarta, J. A. Sea-level stability over geological time owing to limited deep subduction of hydrated mantle. Nature Geosci. 15, 423–428 (2022).
36. Mazza, S.E., Gazel, E., Bizimis, M., Moucha, R., Béguelin, P., Johnson, E.A., McAleer, R.J., Sobolev, A.V. Sampling the volatile-rich transition zone beneath Bermuda. Nature 569, 398–403 (2019).
37. Yang, J., Faccenda, M. Intraplate volcanism originating from upwelling hydrous mantle transition zone. Nature 579, 88–91 (2020).
38. Zhu, X., Balázs, A., Gerya, T., Sun, Z. Secondary plumes formation controlled by interaction of thermochemical mantle plumes with the mantle transition zone. Geophys. Research Let. 52, e2025GL117079 (2025).

**Figure legends**

**Fig. 1. Computed solid-liquid density contrast for MTZ P-T conditions in pirolite mantle ($X_B$=0.17). a**, mantle with 1 wt% water. **b**, mantle with 0.3 wt% water. Coexisting solid and melt compositions are computed with the empirical melting model (Methods). Melt density is computed with the equation of state of Jing and Karato[23]. Solid density is computed with the software PERPLE-X by using the thermodynamic database of Stixrude and Lithgow-Bertelloni[21] (2022) and a water content-related correction (Methods).

**Fig. 2. Development of partially molten mantle plume from solid-state hydrous upwelling at the upper MTZ boundary**. Hydrous upwelling is initially characterized by the negative (-100 K) temperature anomaly and increased water content (0.5 wt%), which causes its positive buoyancy relative to the ambient mantle (0.03 wt% $H_2O$) (model diapir59, Extended Data Table S2). White contour shows the upwelling core. Black lines are isotherms in °C. Blue arrows indicate the velocity field of the solid matrix.

**Fig. 3. Development of small molten diapirs from solid-state hydrous drip at the lower MTZ boundary**. Hydrous downwelling is initially characterized by the negative (-100 K) temperature anomaly, increased water content (0.4 wt%) and enriched composition ($X_B$=0.34), which causes its negative buoyancy relative to the ambient mantle ($X_B$=0.17, 0.03 wt% $H_2O$) (model diapir77, Extended Data Table S2). White contour shows the drip core. Black lines are isotherms in °C. Blue arrows indicate the velocity field of the solid matrix.

**Fig. 4. Development of small molten diapirs from hydrous hot plume at the lower MTZ boundary**. Hydrous plume is initially characterized by the positive (+200 K) temperature anomaly, increased water content (0.2 wt%) and depleted composition ($\frac{X_B}{X_B+X_L}$ = 0.12), which causes its positive buoyancy relative to the ambient mantle ($\frac{X_B}{X_B+X_L}$ = 0.17, 0.03 wt% $H_2O$) (model diapir60, Extended Data Table S2). White contour shows the plume core. Black lines are isotherms in °C. Blue arrows indicate the velocity field of the solid

matrix.

**Fig. 5. Conceptual model of transient water dynamics in the MTZ.** Water is introduced into the MTZ mainly by subducting slabs and (in subordinate amount) by deep hydrous plumes. Three principal types of transient hydrous structures should be characteristic for the heterogeneous MTZ: stagnant fragments of hydrated cold slabs, solid-state hydrous plumes and small melt diapirs.

**Methods**

The following mass conservation equations[4] consistent with Biot's poroelasticity theory and Gassmann's relation[39-41] have been used to account for the volumetric effects of phase transitions (including melting) and compressibility and thermal expansion of melt and solid matrix whereas neglecting compressibility and thermal expansion of pores

*for solid (porous matrix divergence)*

$$div(\vec{v}^s) + \beta_d \left(\frac{D^s P^t}{Dt} - K_{BW}\frac{D^f P^f}{Dt}\right) + \frac{P^t - P^f}{(1-\phi)\eta^\phi} = \alpha^s \frac{D^s T}{Dt} + \Gamma_{mass}\left(\frac{1}{\rho_c^f} - \frac{1}{\rho_c^s}\right), (1)$$

*for fluid (Darcy flux divergence)*

$$div(\vec{q}^D) - K_{BW}\beta_d\left(\frac{D^s P^t}{Dt} - \frac{1}{K_{Sk}}\frac{D^f P^f}{Dt}\right) - \frac{P^t - P^f}{(1-\phi)\eta^\phi} = \phi\left(\alpha^f \frac{D^f T}{Dt} - \alpha^s \frac{D^s T}{Dt}\right), (2)$$

where $\beta_d = \frac{\beta^s + \beta^\phi}{1-\phi}$ is drained compressibility, $K_{BW} = 1 - \frac{\beta^s}{\beta_d}$ is Biot-Willis coefficient, $K_{Sk} = \frac{\beta_d - \beta^s}{\beta_d - \beta^s + \phi(\beta^f - \beta^s)}$ is Skempton's coefficient, $\beta^s$, $\beta^f$ and $\beta^\phi = 0$ are respectively compressibility of solid, fluid and pores ($\beta^\phi$ is assumed to be negligible). $P^t = \phi P^f + (1-\phi)P^s$ is the total pressure, $P^s$ and $P^f$ are respectively solid and fluid pressure, $\phi$ is the porosity (volumetric fluid fraction), $\alpha^s$ is the effective thermal expansion of solid (including effects of solid-state phase transformations), $\alpha^f$ is fluid thermal expansion, $\Gamma_{mass}$ is the rate of mass transfer from solid to fluid due to melting, $\rho_c^s$ and $\rho_c^f$ is the density of the transferred chemical component in the solid and liquid state, respectively[4], $\frac{D^s}{Dt}$ and $\frac{D^f}{Dt}$ denote Lagrangian time derivatives in the solid and fluid velocity frame, respectively, $\vec{v}^s$ is velocity vector of solid, $\vec{q}^D$ is Darcy flux vector, $\eta^\phi = \frac{\eta^s}{\phi}$ is compaction viscosity, $\eta^s$ is shear viscosity of the solid computed with the upper mantle flow law of Karato and Wu[42].

The momentum equations for bulk viscous material and for the fluid are, respectively

$$\frac{\partial \tau_{ij}^t}{\partial x_j} - \frac{\partial P^t}{\partial x_i} = \rho^t g_i, \quad (3)$$

$$q_i^D = -\frac{k^\phi}{\eta^f}\left(\frac{\partial P^f}{\partial x_i} - \rho^f g_i\right), \quad (4)$$

where $\tau_{ij}^t = \eta^t\left(\frac{\partial v_i^s}{\partial x_j} + \frac{\partial v_j^s}{\partial x_i} - \delta_{ij}\frac{1}{3}div(\vec{v}^s)\right)$ is deviatoric stress tensor, $\eta^t = \eta^s e^{-28\phi}$ is shear viscosity of the bulk material[43], $\eta^f$ is shear viscosity of fluid, $k^\phi = k_r^\phi\left(\frac{\phi}{\phi_r}\right)^3\left(\frac{1-\phi_r}{1-\phi}\right)^2$ is the permeability[4,44], $k_r^\phi = 10^{-13}$ m² and $\phi_r = 0.01$ are reference values for permeability and porosity, respectively, $\rho^t = \phi\rho^f + (1-\phi)\rho^s$ is density of bulk material, $\rho^s$ and $\rho^f$ are respectively density of solid and fluid, $x_i$ are coordinates, $\delta_{ij}$ is the Kronecker delta.

The energy conservation equation is formulated for the bulk material as[4]

$$\rho^t C_P^t\left(\phi\frac{D^f T}{Dt} + (1-\phi)\frac{D^s T}{Dt}\right) = -\frac{\partial q_i^t}{\partial x_i} + H_r^t + H_a^t + H_s^t + H_L^t, \quad (5)$$

$$q_i^t = -k^t\frac{\partial T}{\partial x_i}, \quad (6)$$

$$H_r^t = (1-\phi)H_r^s + \phi H_r^f, \quad (7)$$

$$H_a^t = \left(1 - \rho^t\frac{\partial H^t}{\partial P}\right)\frac{DP^t}{Dt}, \quad (8)$$

$$H_s^t = \frac{(\tau_{ij}^t)^2}{2\eta^t} + \frac{(P^t-P^f)^2}{(1-\phi)\eta^\phi} + \frac{\eta^f}{k^\phi}(q_i^D)^2, \quad (9)$$

$$H_L^t = -\Delta H^{f-s}\Gamma_{mass}, \quad (8)$$

where $C_P^t$ is effective isobaric heat capacity of bulk material (including effects of solid-state phase transformations), $q_i^t$ are components of bulk heat flux vector, $H_r^t, H_a^t, H_s^t$ and $H_L^t$ is volumetric radioactive, adiabatic, shear and latent heat production, respectively, $\Delta H^{f-s}$ is the enthalpy difference between the fluid and the solid, $\frac{\partial H^t}{\partial P}$ is pressure derivative of enthalpy of bulk material (including effects of solid-state phase transformations), $k^t = \sqrt{\frac{k^s k^f}{2} + \frac{[k^s(3\phi-2)+k^f(1-3\phi)]^2}{16}} -$ $\frac{k^s(3\phi-2)+k^f(1-3\phi)}{4}$ is thermal conductivity of bulk material[4,45], $k^s = 3$ W/m/K and $k^f = 5$ W/m/K are respectively thermal conductivity of solid and fluid.

Mass transfer-related terms linked to melting/solidification are computed numerically[4] on the basis of the new MTZ melting model

$$\Gamma_{mass}\left(\frac{1}{\rho_c^f} - \frac{1}{\rho_c^s}\right) = \frac{1-R_V}{\Delta t},$$

$$\Gamma_{mass} = \frac{\rho^{f\Delta t}\phi^{\Delta t} - R_V \rho^{fo}\phi^o}{\Delta t},$$

where $R_V = \left[\frac{\rho^{so}(1-\phi^o)+\rho^{fo}\phi^o}{\rho^{s\Delta t}(1-\phi^{\Delta t})+\rho^{f\Delta t}\phi^{\Delta t}}\right]^{-1}$ is the initial (i.e., for the beginning of the time step) to final (i.e., for the end of the time step) local system volume ratio for the given time step, $\Delta t$ is the time step size, $\phi^o, \rho^{so}, \rho^{fo}$ and , $\phi^{\Delta t}, \rho^{s\Delta t}, \rho^{f\Delta t}$ are porosity and density of the solid and fluid for the beginning and the end of the time step, respectively.

**Empirical model of MTZ melting**

We developed a new empirical model of deep hydrous mantle melting at characteristic MTZ P-T conditions of 10-25 GPa, which is based on available experimental data for both dry and hydrous conditions[16-20]. In order to account in a simplified manner for compositional and density variations of coexisting melt and solid, we followed the approach presented by Gerya[4] and considered three end-members that can mix in different proportions in liquid and solid phase: depleted silicate component (*L*), enriched silicate component (*B*) and water (*H*). When calibrating compositions of the two silicate components we mainly focused on reproducing the experimentally determined variations of H$_2$O content and FeO/(FeO+MgO+CaO) and (SiO$_2$+Al$_2$O$_3$)/(SiO$_2$+Al$_2$O$_3$+MgO+CaO) ratios in the liquid, which are responsible for the largest variations in the melt density[20,23]. Our empirical melting model assumes ideal mixing of all three melt components in both solid and liquid state. This empirical model is obviously very simplified and neglects actual mineralogy of the solid matrix coexisting with the melt focussing instead on reproducing the experimentally determined degree of mantle melting and density variations of the resulting liquid in the P-T range characteristic for the MTZ. Three end-member reactions then characterise the melting process melting of the enriched silicate component $B$:   $B^s = B^f$, melting of the depleted silicate component $L$:   $L^s = L^f$, dehydration of wadsleyite:   $H^{wad} = H^f$, where superscripts $s$, $f$ and $wad$ stand for solid, liquid (aqueous fluid, melt) and wadsleyite, respectively.

Thermodynamic equilibrium conditions for these three reactions are formulated as

$$\Delta G_{B(P^f,T)} = 0 = \Delta H_B - T\Delta S_B + P^f \Delta V_B + RT\ln\left(\frac{X_B^f}{X_B^s}\right),$$

$$\Delta G_{L(P^f,T)} = 0 = \Delta H_L - T\Delta S_L + P^f \Delta V_L + RT\ln\left(\frac{X_L^f}{X_L^s}\right),$$

$$\Delta G_{H(P^f,T)} = 0 = \Delta H_H - T\Delta S_H + P^f \Delta V_H + RT\ln\left(\frac{X_H^f}{X_H^{wad}}\right)$$

$$X_B^f + X_L^f + X_H^f = 1,$$

$$X_B^s + X_L^s + X_H^s = 1,$$

$$X_H^s = X_H^{wad} X_{wad}^s + X_H^{ol} X_{ol}^s + X_H^{ring} X_{ring}^s + X_H^{pv} X_{pv}^s,$$

$X_B^s, X_L^s, X_H^s$ and $X_B^f, X_L^f, X_H^f$ are molar fractions of respective components in the solid and melt, respectively, $X_H^{wad}, X_H^{ol}, X_H^{ring}, X_H^{pv}$ are molar fractions of water in respectively wadsleyite, olivine, ringwoodite, perovskite coexisting with liquid, $X_{wad}^s, X_{ol}^s, X_{ring}^s, X_{pv}^s$ are molar fractions of respective minerals in solid, $R$=8.314 J/K/mol is the gas constant, $\Delta G$, $\Delta H$, $\Delta S$ and $\Delta V$ are molar Gibbs free energy, enthalpy, entropy and volume change in respective reactions. Mass balance constraints are given by the composition of the system $X_B^t, X_L^t, X_H^t$

$$X_B^t = X_B^s X^s + X_B^f X^f$$

$$X_L^t = X_L^s X^s + X_L^f X^f$$

$$X_H^t = X_H^s X^s + X_H^f X^f$$

$$X_B^t + X_L^t + X_H^t = 1$$

$$X^s + X^f = 1$$

where $X^s$ and $X^f$ are equilibrium molar fractions of solid and fluid in the system.

Water content in wadsleyite ($X_H^{wad}$) has been computed from its experimentally determined stoichiometry[18]. Partition coefficients between wadsleyite and other water-bearing nominally anhydrous minerals (NAMs) are taken from Inoue et al.[22]. Calibrated parameters for the three end member reactions are given in Extended Data Table S1. Dry pyrolite solidus and liquidus for this empirical model are shown in Extended Data Fig.1a.

**Density and other thermodynamic properties**

Our melting model is able to predict chemical compositions of coexisting melt and solid in the CaO-FeO-MgO-Al$_2$O$_3$-SiO$_2$-H$_2$O (CFMASH) system with compositional variations along the ternary B-L-H mixing trend. In order to compute liquid density in this system we used equation of state of Jing and Karato[23]. Dry solid mineralogical composition (including $X_H^{wad}, X_H^{ol}, X_H^{ring}, X_H^{pv}$), density, compressibility has been computed with the PERPLE-X Gibbs free energy minimization software by using the thermodynamic database of Stixrude and Lithgow-Bertelloni[21]. Water incorporation into wadsleyite, olivine, ringwoodite and perovskite is computed on the basis of the calibrated wadsleyite dehydration reaction and partition coefficients between wadsleyite and other three NAMs (olivine, ringwoodite and perovskite)[22]. Density changes of the dry solid related to water incorporation are computed on the basis of pure water density predicted by the equation of state of Jing and Karato[23] and the calibrated negative volumetric effect of wadsleyite dehydration reaction (Extended Data Table S1). Effective $C_P^t$ and $\frac{\partial H^t}{\partial P}$ of the bulk dry material (including effects of solid-state phase transitions) are computed for the given bulk composition from the thermodynamic database of Stixrude and Lithgow-Bertelloni[21] with the software PERPLE-X. Water-related correction of these parameters is based on the Gibbs free energy equation for water of Gerya et al.[24], which works well for MTZ P-T conditions and gives water density values consistent with the equation of state of Jing and Karato[23]. In order to speedup calculations, all thermodynamic properties were precomputed in form of lookup tables in $P - T - \frac{X_B}{X_B+X_L} - X_H$ space. Comparison of melt density for experimentally measured and computed equilibrium melt compositions in both dry and hydrous system is shown in Extended Data Fig.1b.

**References of Methods**

39. Biot, M. A. General theory of three-dimensional consolidation. J. Applied Phys. 12, 155–164 (1941).
40. Gassmann, F. Über die elastizität poröser medien. Vierteljahrsschrift der Naturforschenden Gesellschaft in Zürich, 96, 1–23 (1951).
41. Yarushina, V. M., Podladchikov, Y. Y. (De)compaction of porous viscoelastoplastic media: model formulation. J. Geophys. Res. 120, 4146–4170 (2015).
42. Karato, S., Wu, P. Rheology of the upper mantle: a synthesis. Science 260, 771–778 (1993).
43. Katz, R. F., Spiegelman, M., Holtzman, B. The dynamics of melt and shear localization in partially molten aggregates. Nature, 442, 676–679 (2006).
44. Yarushina, V. M., Bercovici, D., Oristaglio, M. L. Rock deformation models and fluid leak-off in hydraulic fracturing. Geophysical Journal International, 194, 1514–1526 (2013).
45. Budiansky, B. Thermal and thermoelastic properties of isotropic composites. Journal of Composite Materials, 4, 286–295 (1970).

**Data availability**

All data files used for the empirical melting model calibration and for running numerical experiments will be made publicly available upon publication of the paper.

**Code availability**

The MatLab codes used for numerical experiments and visualization will be made publicly available upon publication of the paper.

**Acknowledgements** This work was supported by SNSF research projects 200021-231594 and 200021_192296 (to T.V.G.) by SNSF Sinergia grant N° 213539 (to N.M.B and T.V.G.) and by SNSF research visit project IZSEZ0_230945 (to S.K). Zhicheng Jing is thanked for providing the MatLab code for melt density calculation. The simulations were performed with the ETH-Zurich Euler cluster.

**Extended Data figure legends**

**Extended Data Fig. 1. a, Experimental data used for calibration of the empirical melting model. b, Comparison of modeled and experimental melt densities.** Experimental data are taken from literature: L&O_2002[18], MEA_2006[19], I&S_1992[16], FEA_2025[20], Z&H_1994[17]. Both experimental and modeled densities in b are computed from EOS of Jing and Karato[23] based on experimentally measured and computed melt composition, respectively. With exception of experiments of Fei et al.[20] who provided measured water content in the melt, water content in hydrous experimental melts is computed from the empirical melting model developed in this work. C(F)MAS(H) corresponds to $CaO-(FeO)-MgO-Al_2O_3-SiO_2-(H_2O)$ system.

**Extended Data Fig. 2. Computed solid-liquid density contrast for MTZ P-T conditions in enriched mantle ($X_B$=0.34). a**, mantle with 1 wt% water. **b**, mantle with 0.3 wt% water. Coexisting solid and melt compositions are computed with the empirical melting model (Methods). Melt density is computed with the equation of state of Jing and Karato[23]. Solid density is computed with the software PERPLE-X by using thermodynamic database of Stixrude and Lithgow-Bertelloni[21] and water content-related correction (Methods). Note color scale difference with Fig.1.

**Extended Data Fig. 3. Development of porosity waives from solid-state hydrous upwelling at the upper MTZ boundary**. Hydrous upwelling is initially characterized by the negative (-100 K) temperature anomaly and increased water content (0.7 wt% $H_2O$), which causes its positive buoyancy relative to the ambient mantle (0.2 wt% $H_2O$) (model diapir75, Extended Data Table S2). White contour shows the plume core. Black lines are isotherms in °C. Blue arrows indicate the velocity field of the solid matrix. Note faster movement of porosity waives compared to the plume core.

**Extended Data Fig. 4. Development of Rayleigh-Taylor instability from homogeneously hydrated MTZ**. MTZ mantle is initially characterized by an increased water content (0.4 wt%), which is higher compared to the ambient mantle (0.03 wt% $H_2O$). Small Gaussian thermal perturbation of 10 K is prescribed in the center of the model at the depth of 405 km, which triggers the instability (model diapir62, Extended

Data Table S2). Note apparently stable initial density stratification in **a**. White line shows the boundary between the water-rich MTZ mantle and water-poor upper mantle. Black lines are isotherms in °C. Blue arrows indicate the velocity field of the solid matrix.

**Extended Data Fig. 5. Penetration of hydrous depleted hot plume through the lower MTZ boundary**. Hydrous plume is initially characterized by the positive (+200 K) temperature anomaly, increased water content (0.2 wt%) and depleted composition ($\frac{X_B}{X_B+X_L} = 0.12$), which causes its positive buoyancy relative to the ambient mantle ($\frac{X_B}{X_B+X_L} = 0.17$, 0.03 wt% $H_2O$) (model diapir80, Extended Data Table S2). White contour shows the plume core. Black lines are isotherms in °C. Blue arrows indicate the velocity field of the solid matrix.

**Extended Data Fig. 6. Development of hydrous plumes and melt diapirs from cold hydrated slab fragment located above the lower MTZ boundary**. Slab fragment is initially characterized by the strong negative (-500 K) temperature anomaly and increased water content (0.4 wt% $H_2O$), which causes its initial negative buoyancy relative to the ambient mantle (0.03 wt% $H_2O$) and partial penetration through the lower MTZ boundary (model diapir67, Extended Data Table S2). White contour shows the fragment core. Black lines are isotherms in °C. Blue arrows indicate the velocity field of the solid matrix. Note development of solid-state hydrous plumes and small melt diapirs inside the MTZ.

**Extended Data Fig. 7. Development of hydrous plume from cold hydrated slab fragment located above the lower MTZ boundary**. Slab fragment is initially characterized by the moderate negative (-100 K) temperature anomaly and increased water content (0.4 wt% $H_2O$), which causes its initial positive buoyancy relative to the ambient mantle (0.03 wt% $H_2O$) (model diapir74, Extended Data Table S2). White contour shows the fragment core. Black lines are isotherms in °C. Blue arrows indicate the velocity field of the solid matrix. Note development of solid-state hydrous plumes and small melt diapirs inside the MTZ.

**Extended Data Fig. 8. Development of partially molten mantle plume and**

**descending small melt diapirs from solid-state hydrous upwelling at the upper MTZ boundary under condition of negatively buoyant hydrous melt**. Hydrous upwelling is initially characterized by the negative (-100 K) temperature anomaly and increased water content (0.5 wt%), which causes its positive buoyancy relative to the ambient mantle (0.03 wt% $H_2O$) (model diapir104, Extended Data Table S2). Melt density is increased by adding an ad hock value of 400 kg/m$^3$ to the computed melt density. White contour shows the upwelling core. Black lines are isotherms in °C. Blue arrows indicate the velocity field of the solid matrix.

**Extended Data Table 1.** Calibrated parameters of the empirical melting model used in numerical experiments.

**Extended Data Table 2.** Conditions and results of numerical experiments.

Figure 1

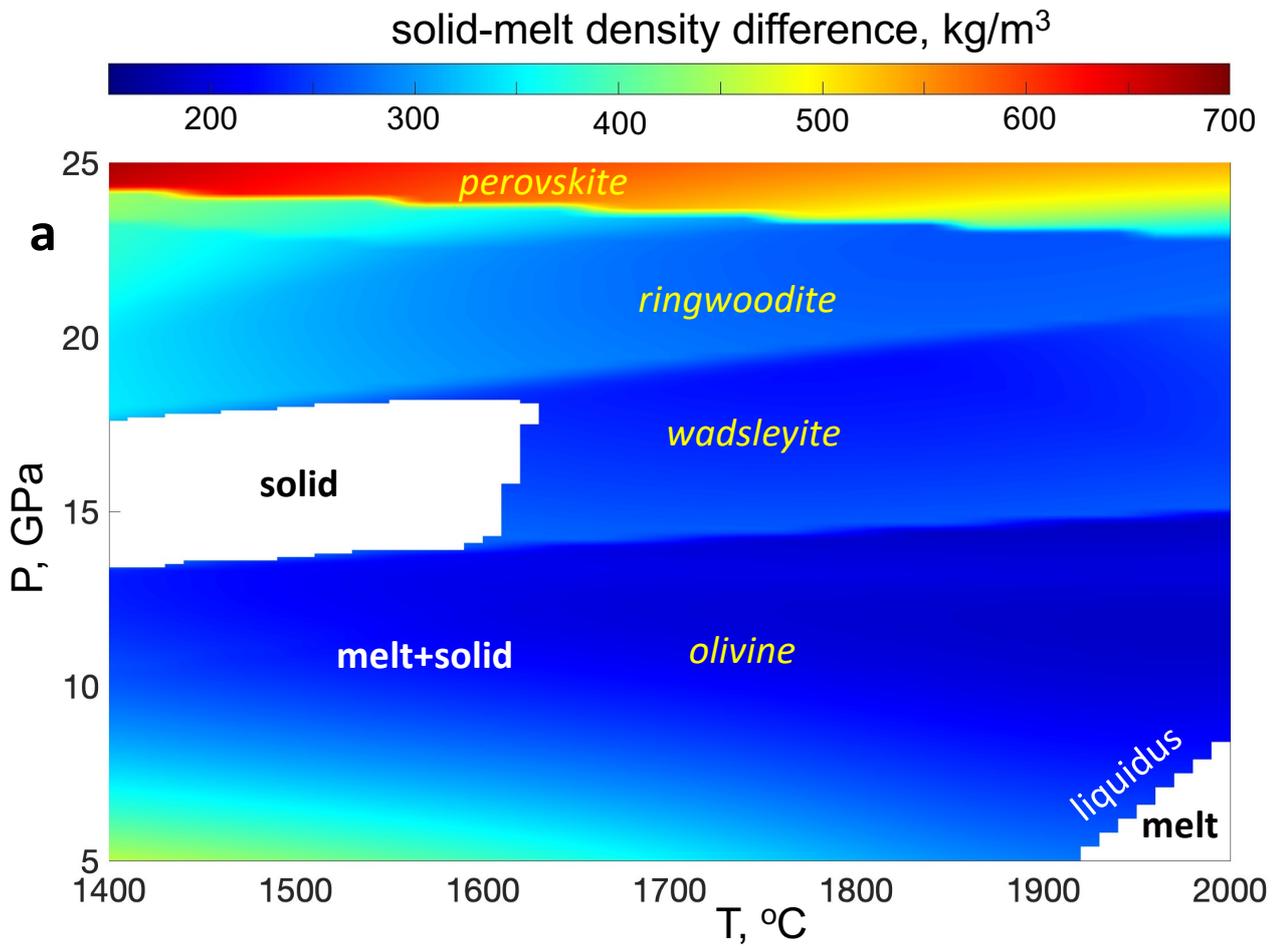

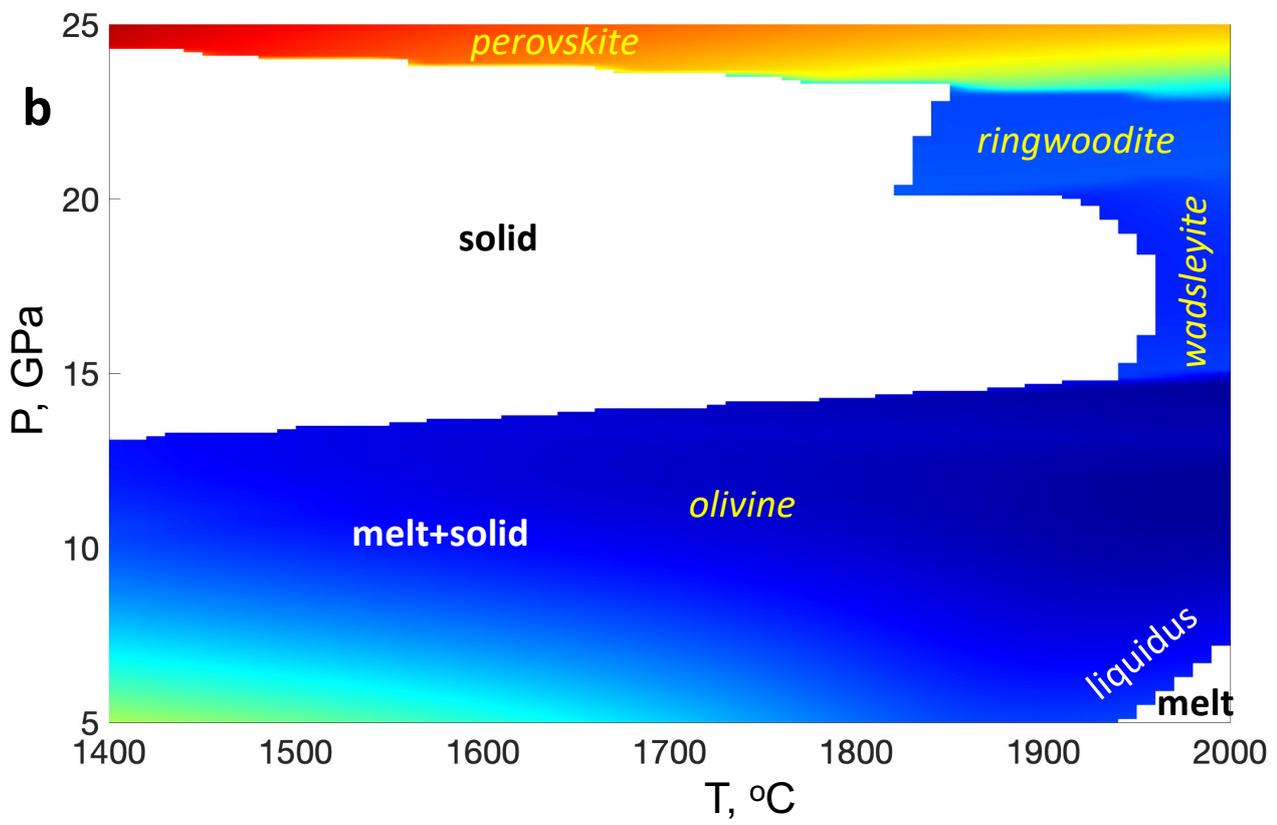

Figure 2

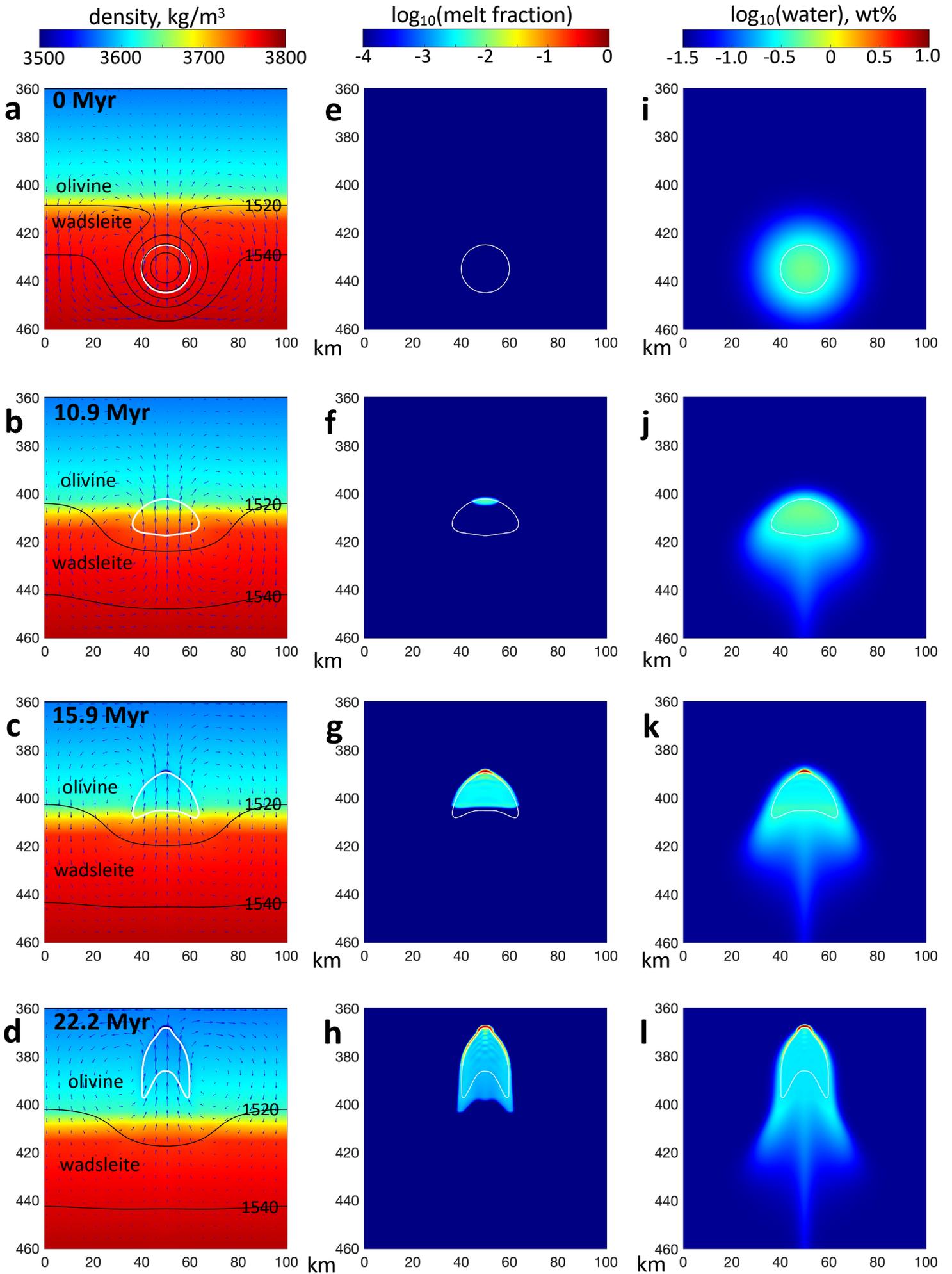

Figure 3

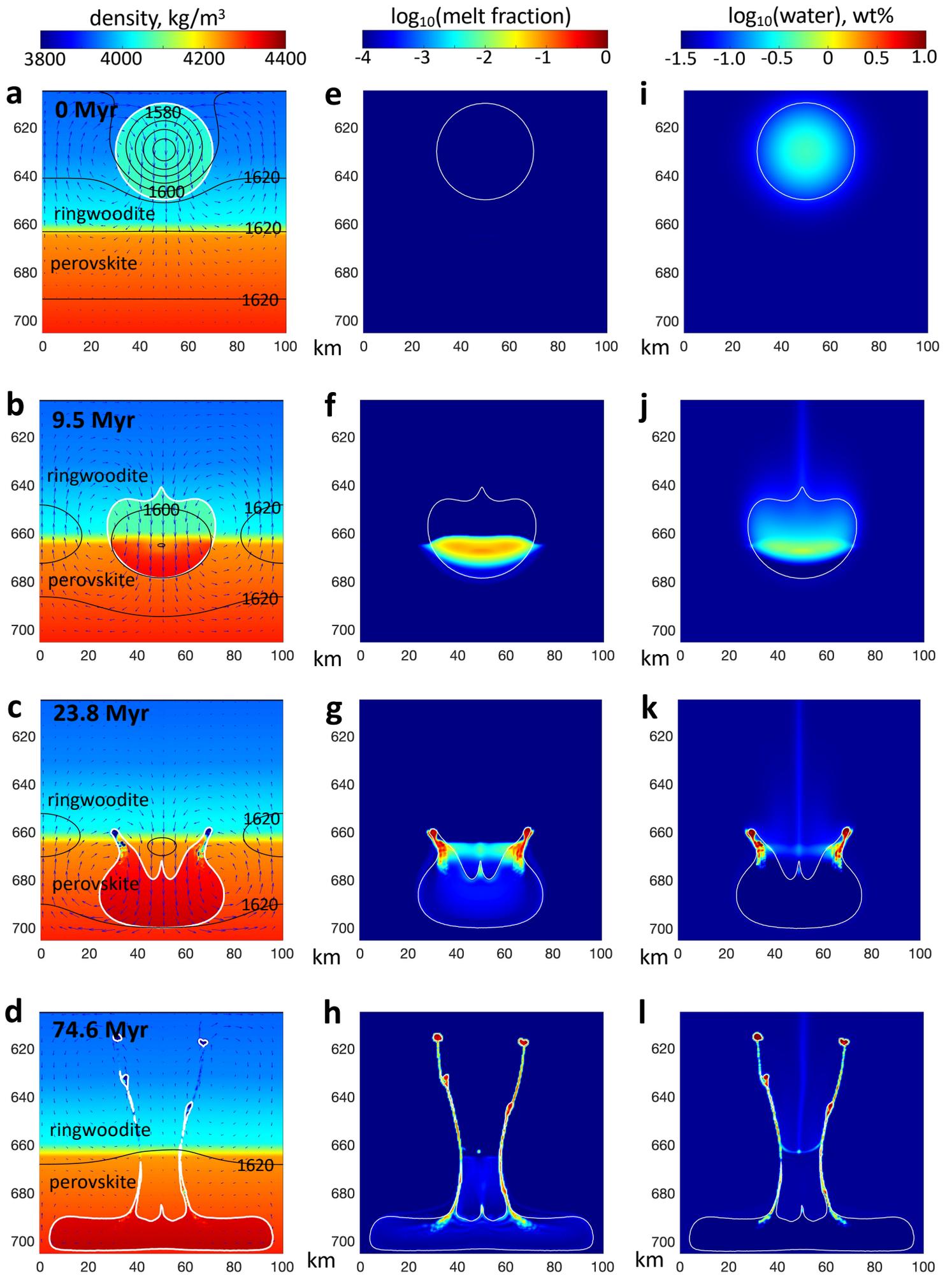

Figure 4

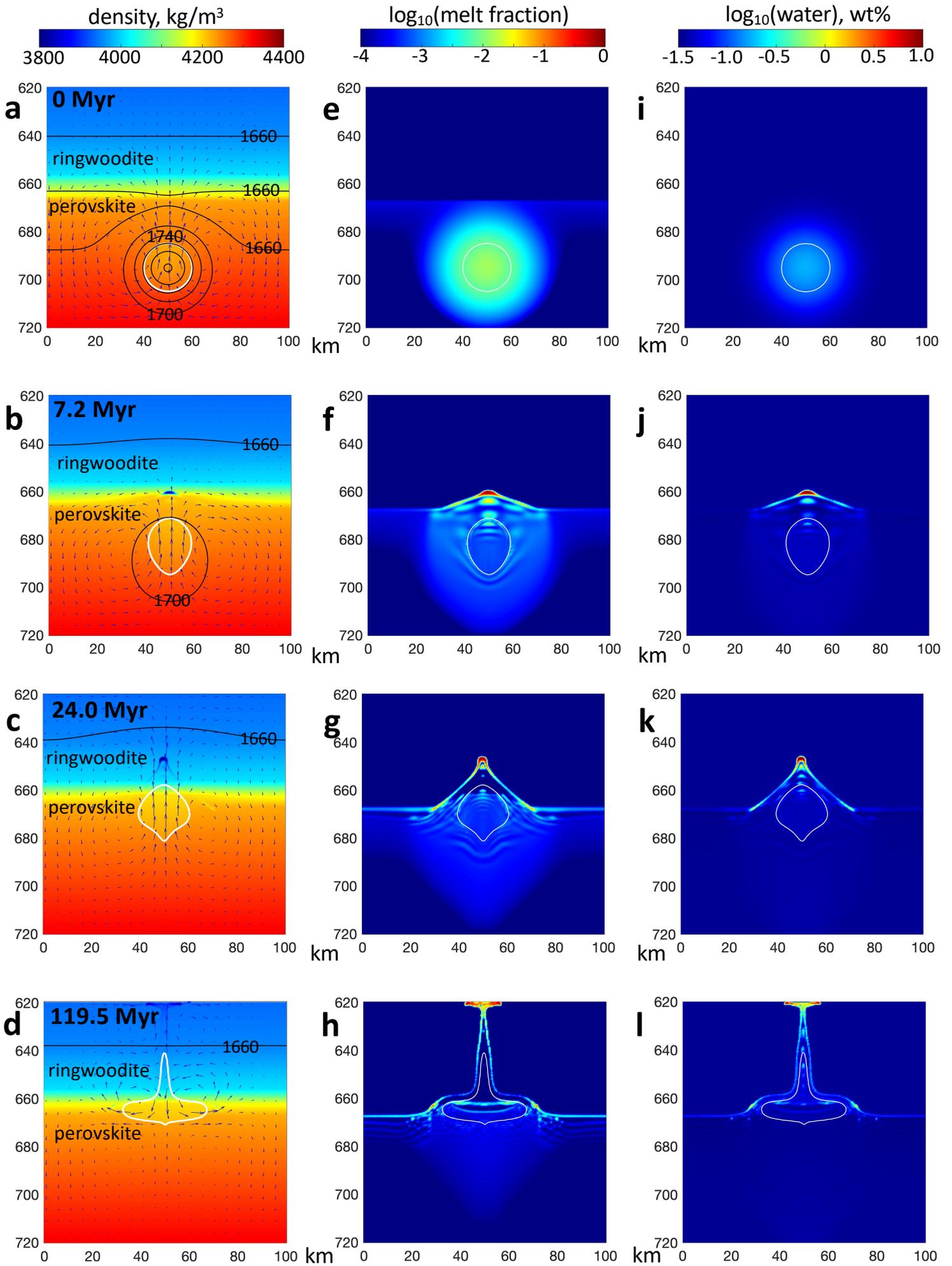

Figure 5

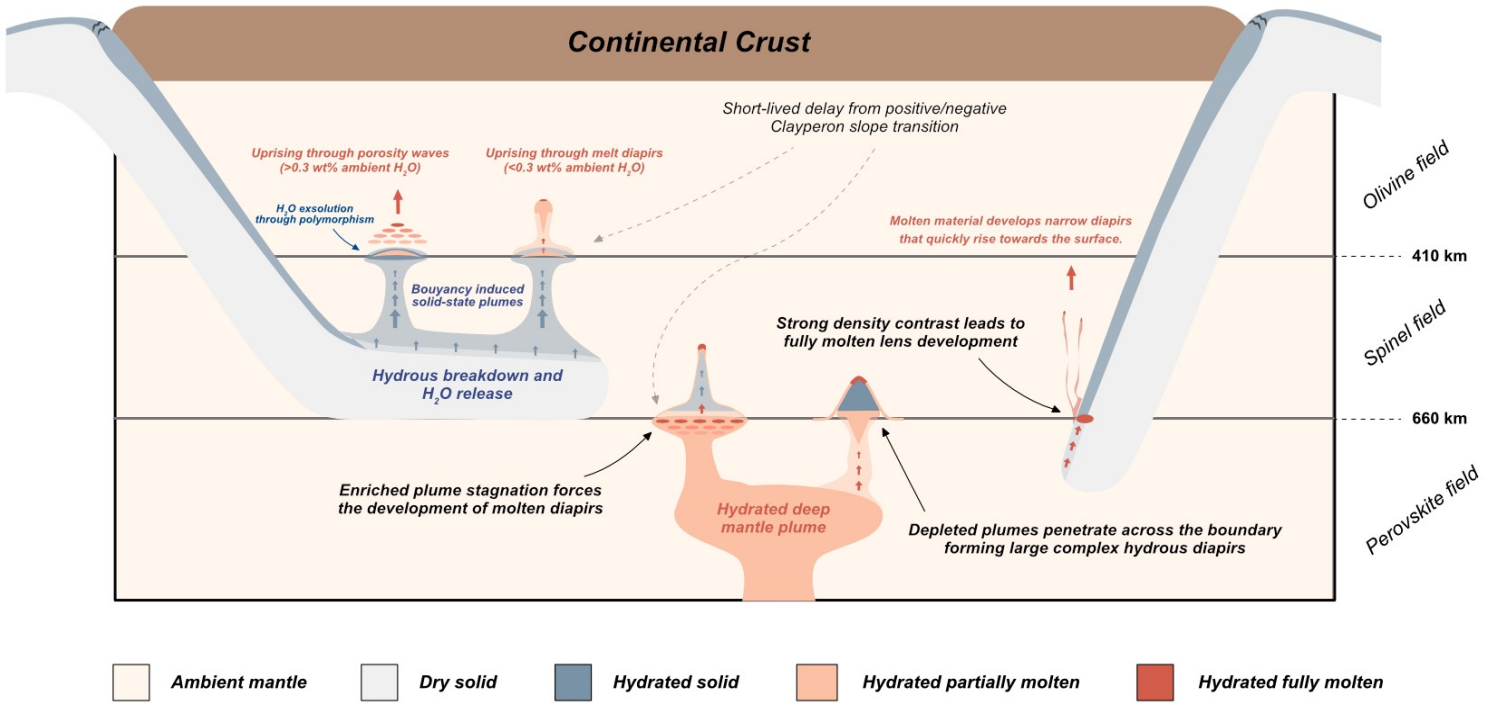

Extended Data Figure 1

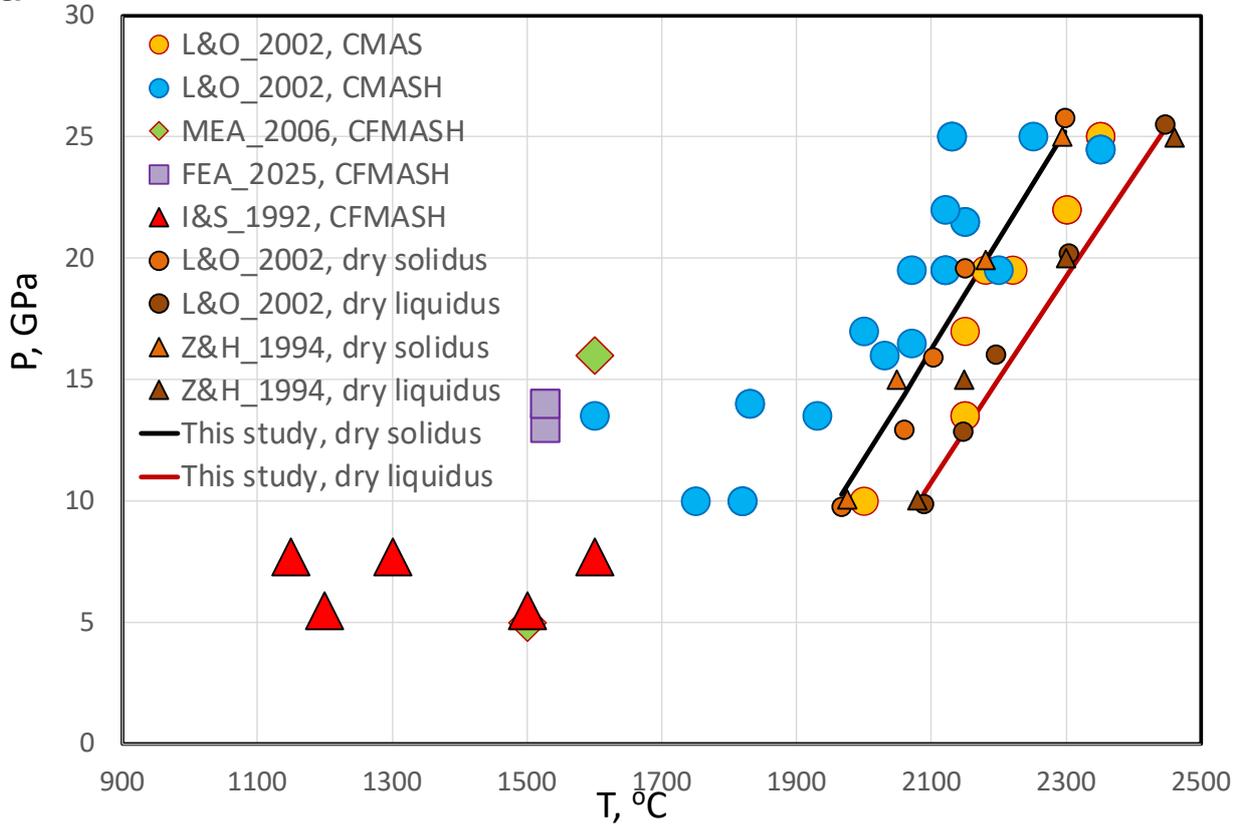
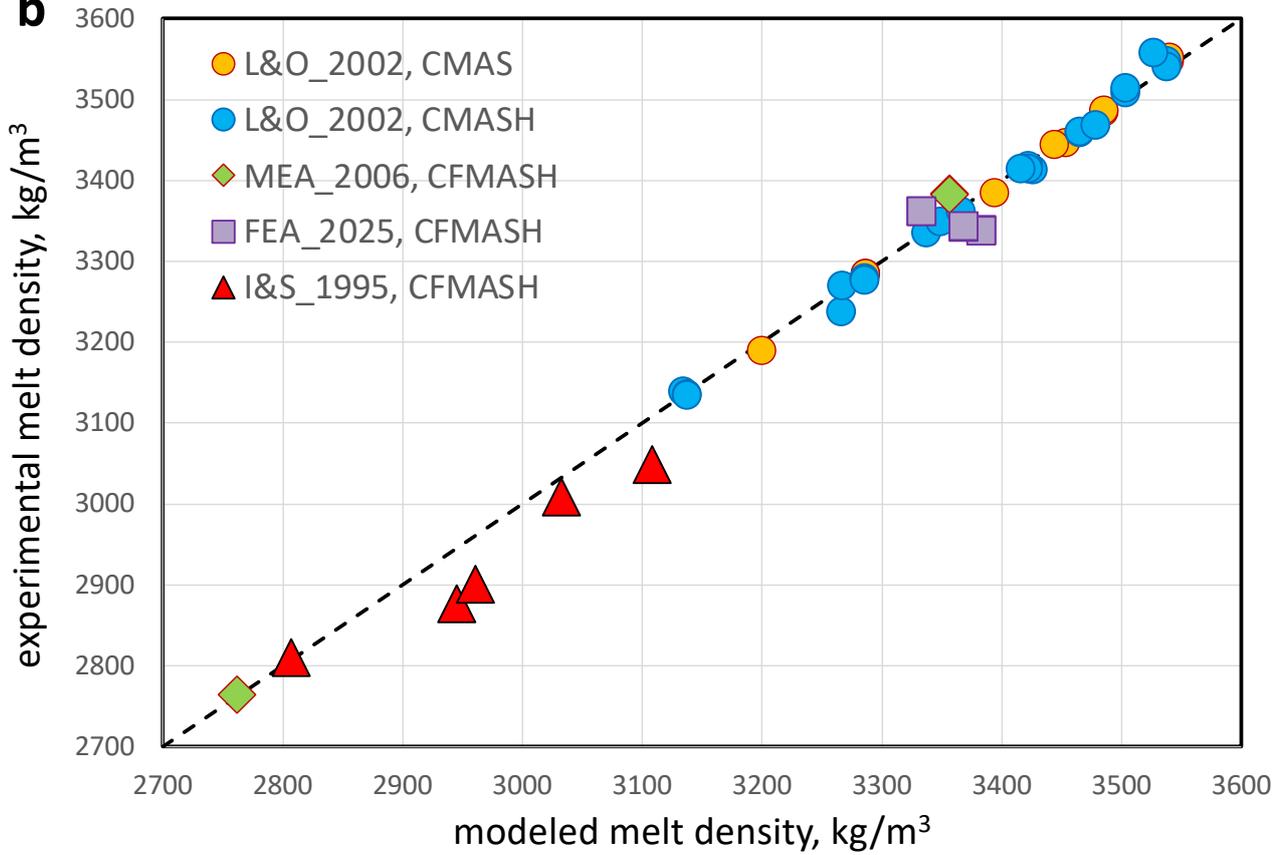

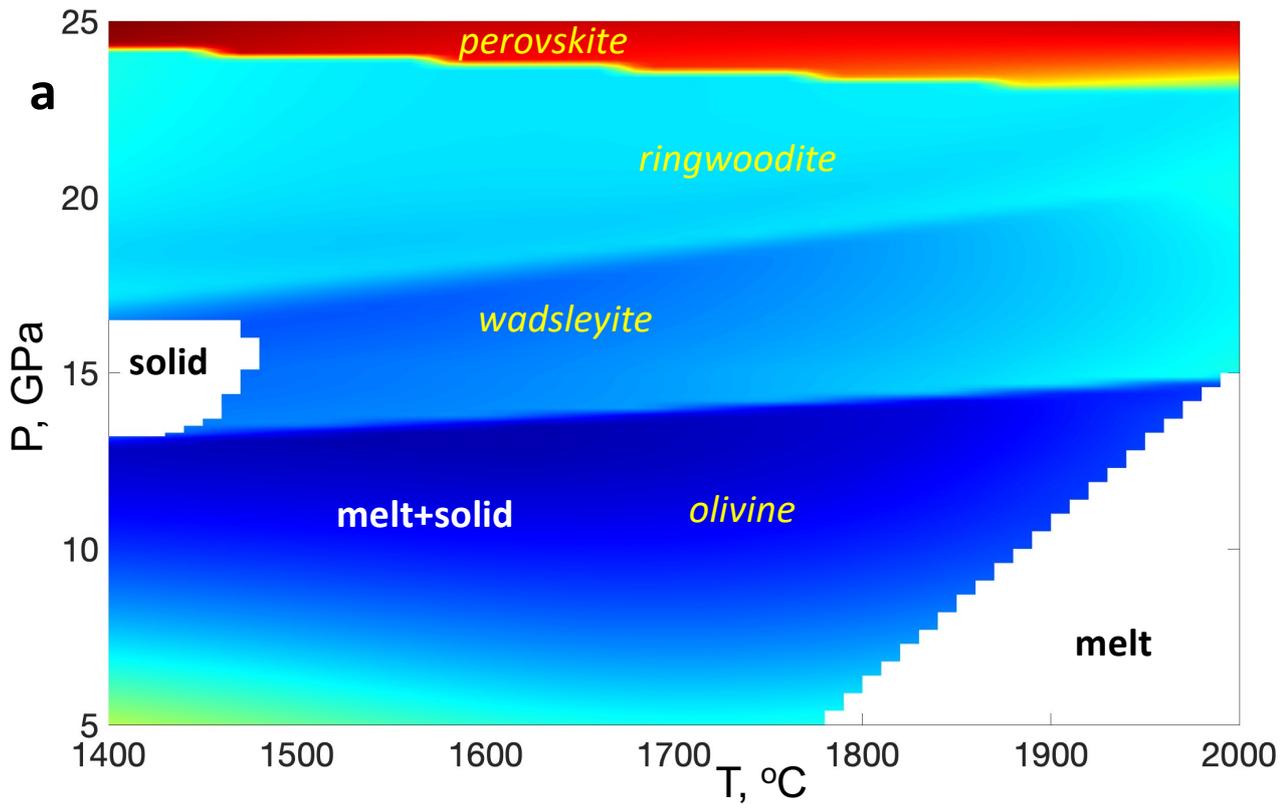
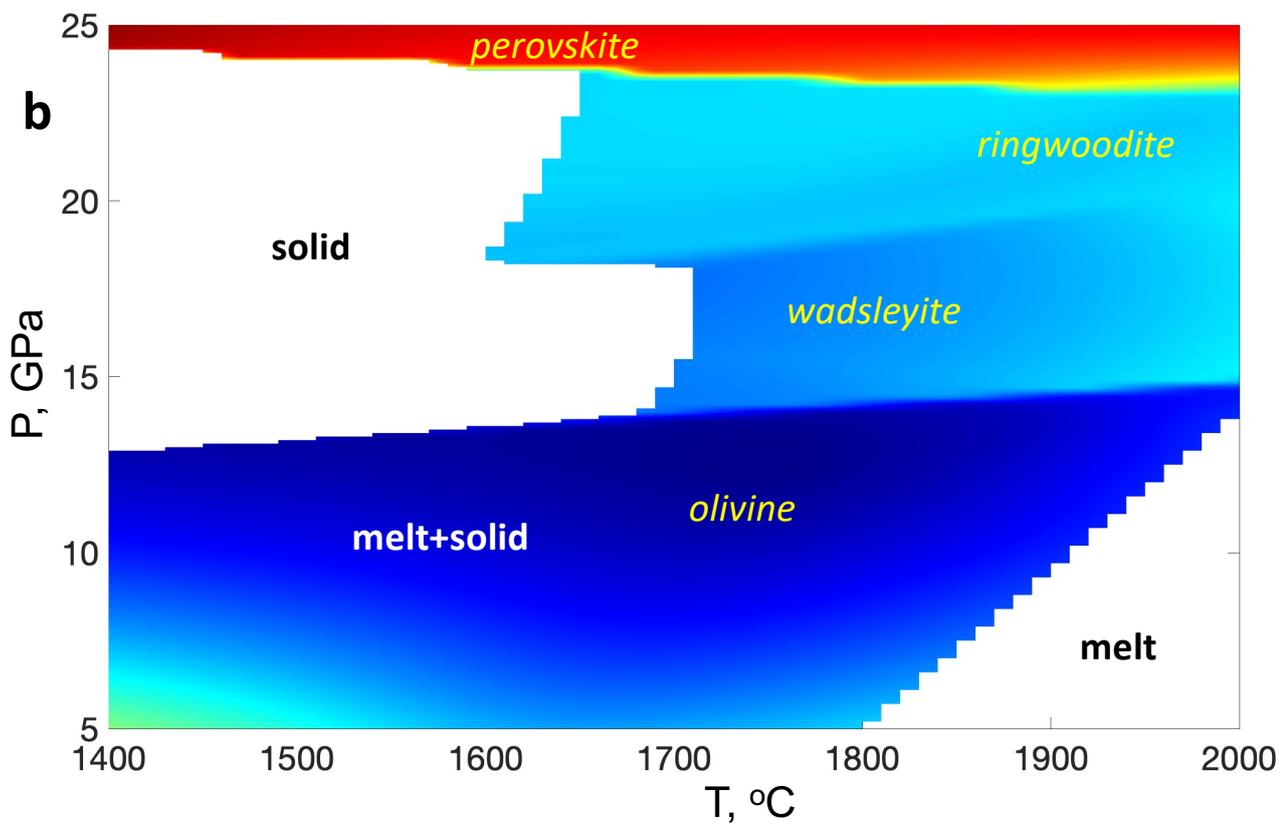

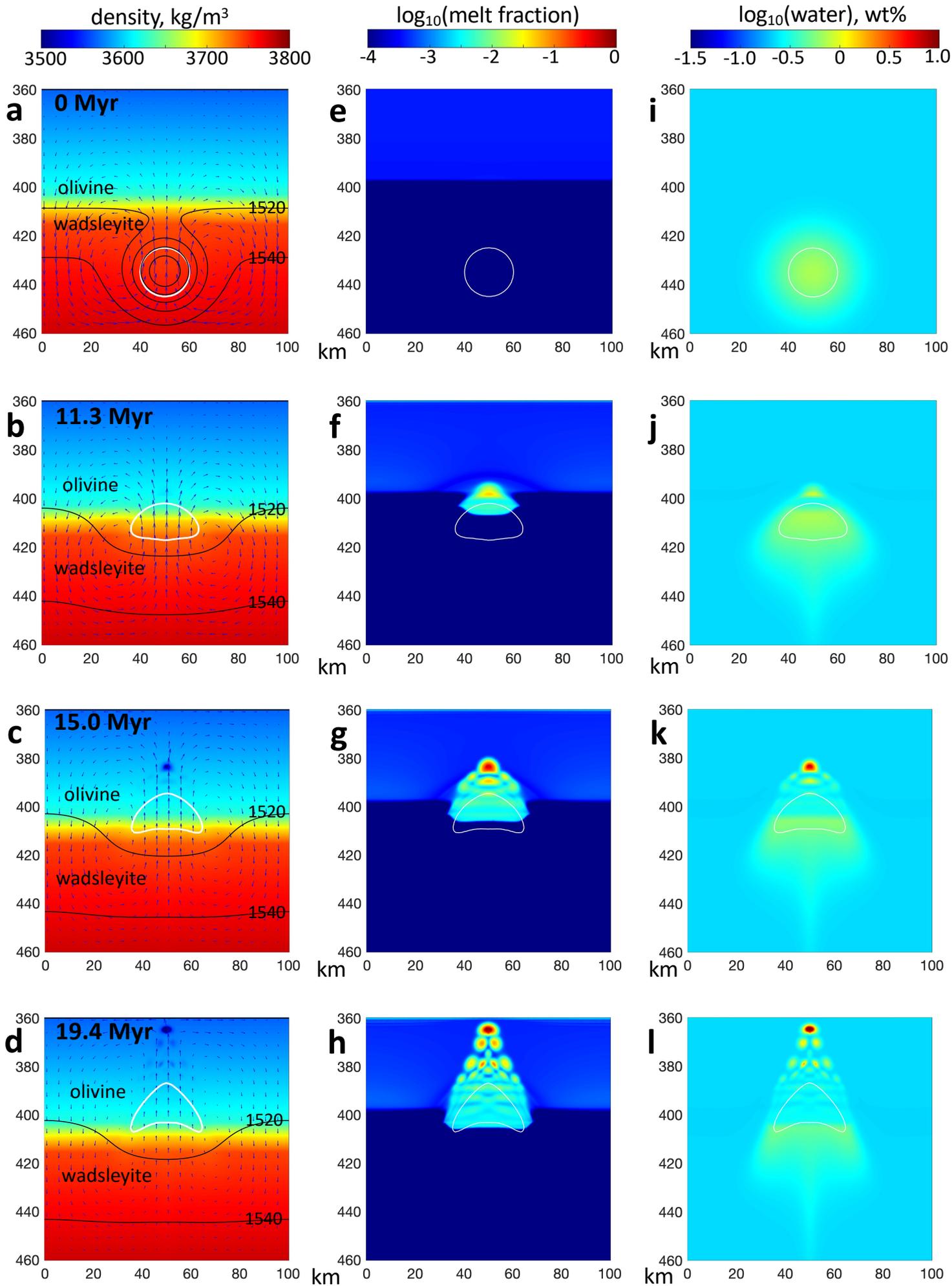

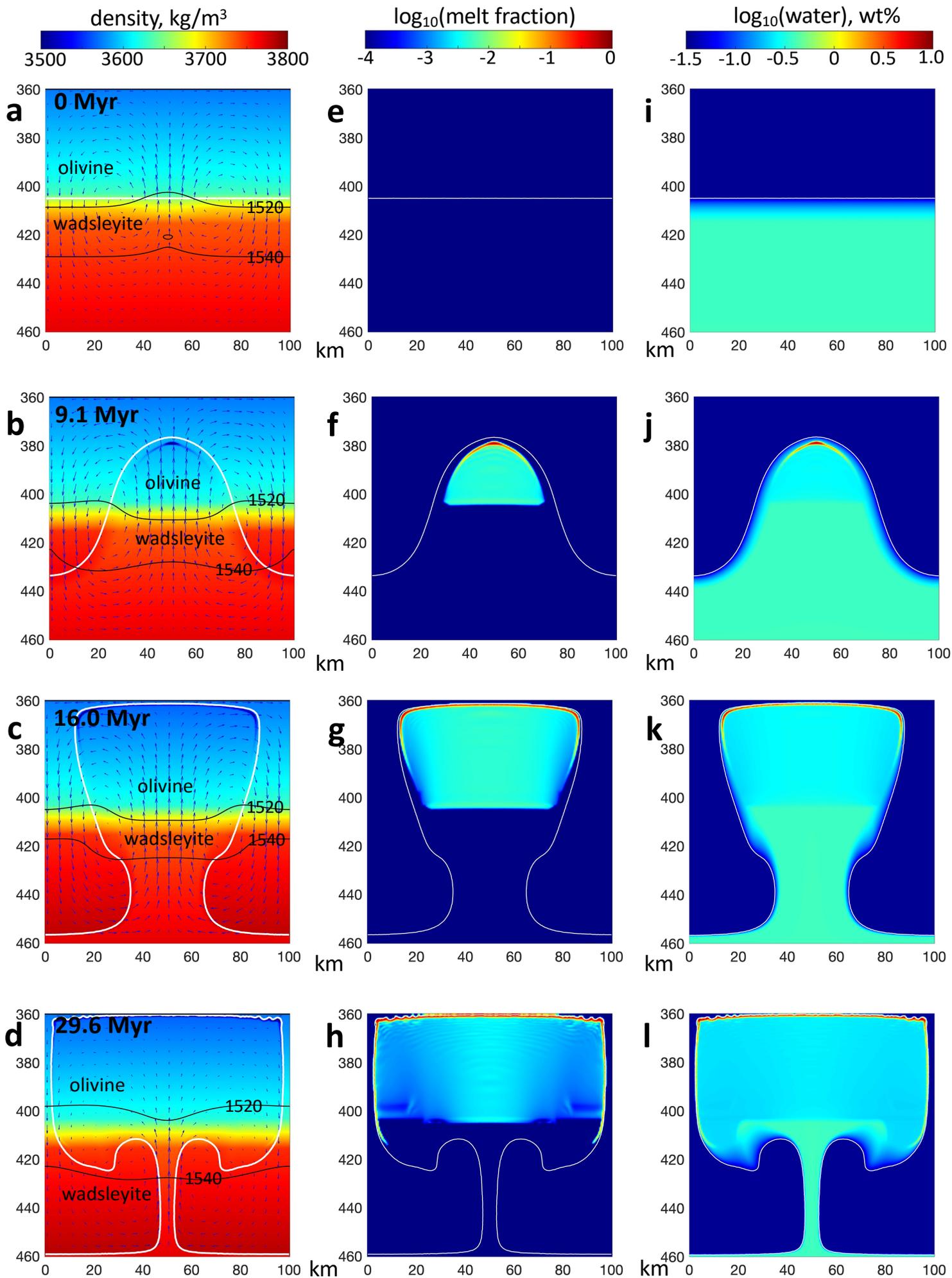

Extended Data Figure 4

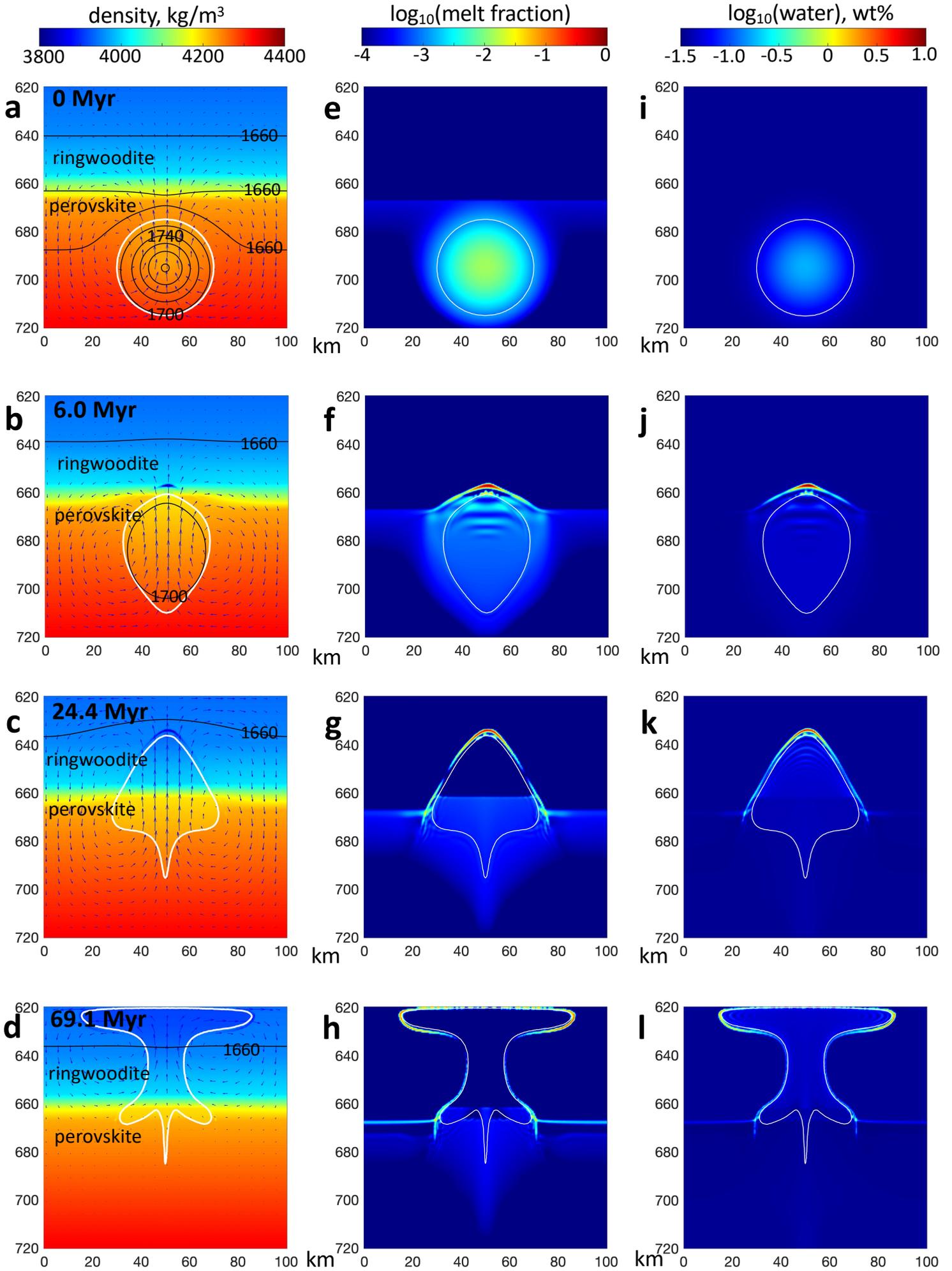

Extended Data Figure 5

Extended Data Figure 6

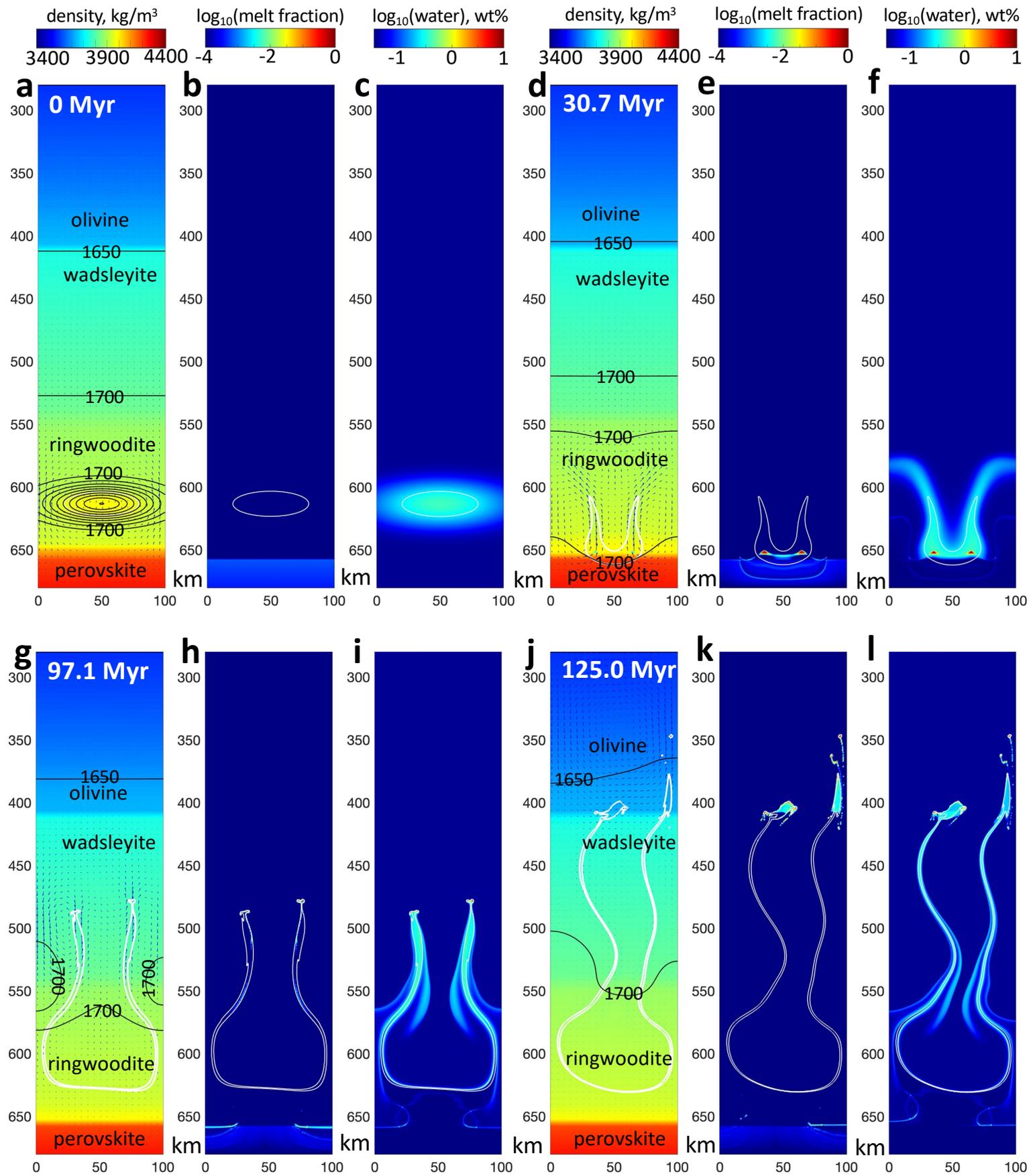

Extended Data Figure 7

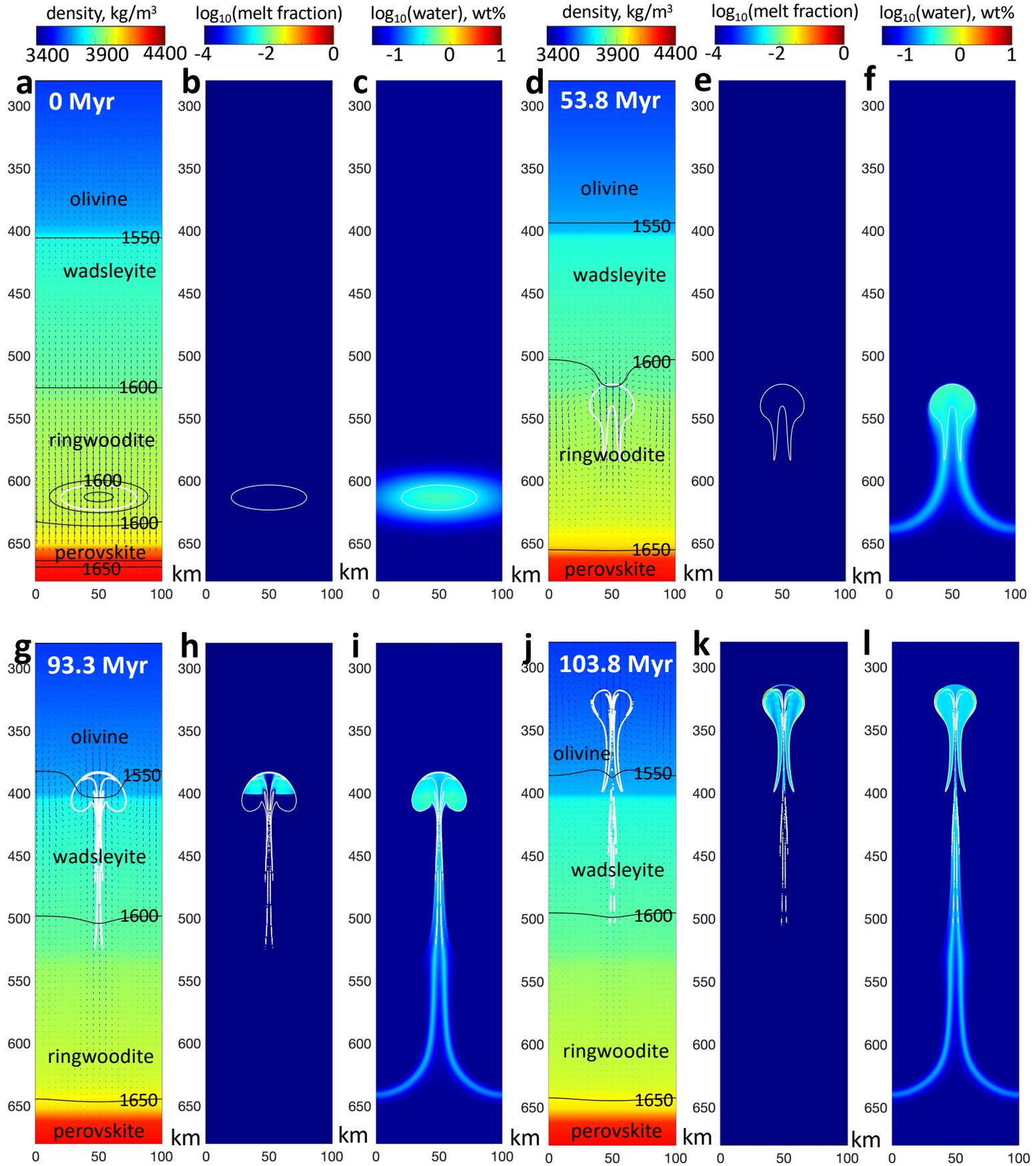

Extended Data Figure 8

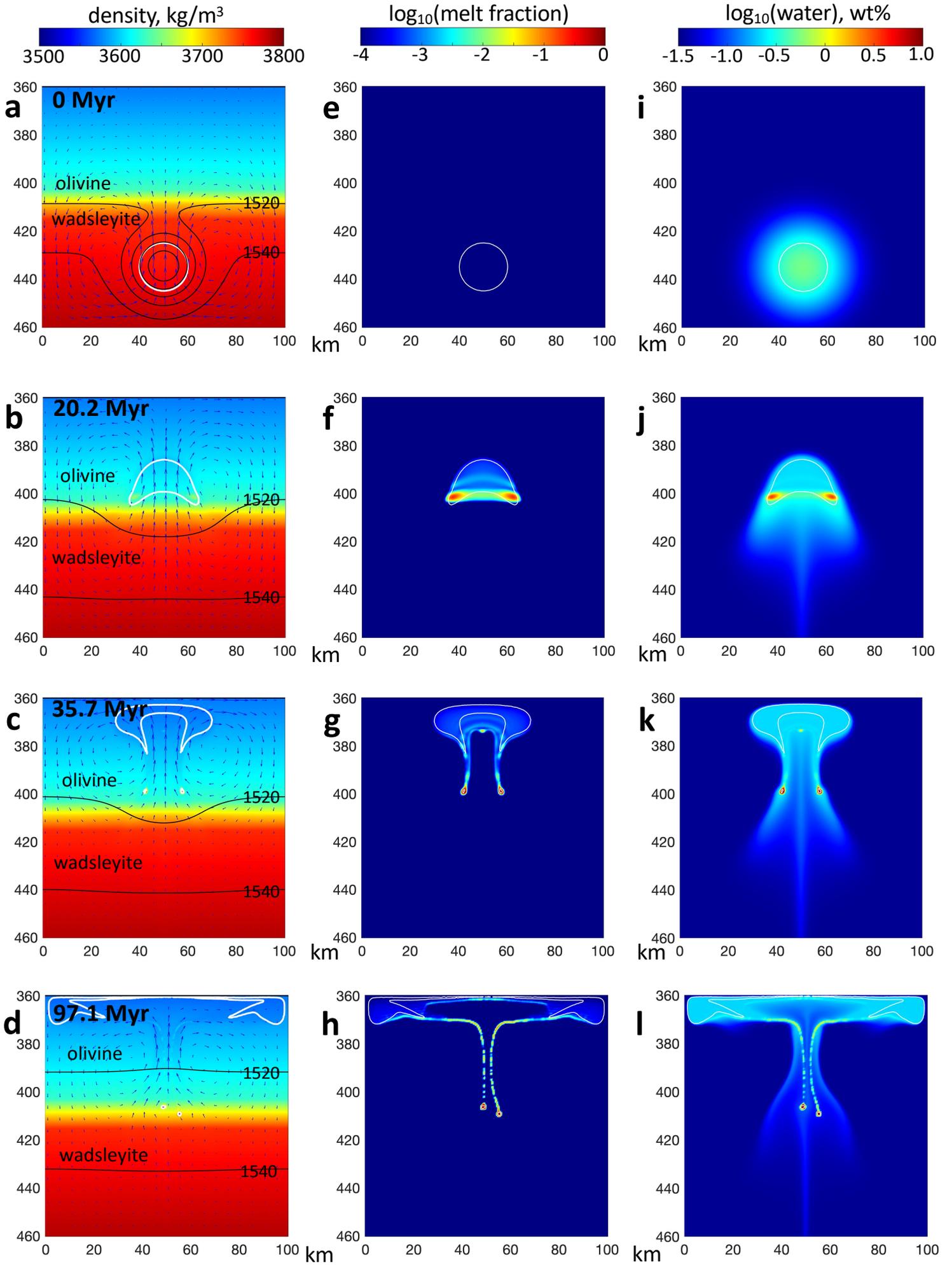

**Extended Data Table 1.** Calibrated parameters of the empirical melting model used in numerical experiments.

| Parameter | Units | Value |
| --- | --- | --- |
| $\Delta H_B$ | J/mol | 13740 |
| $\Delta S_B$ | J/K/mol | 11.635 |
| $\Delta V_B$ | J/bar/mol | 0.013748 |
| $\Delta H_L$ | J/mol | 27797 |
| $\Delta S_L$ | J/K/mol | 12.399 |
| $\Delta V_L$ | J/bar/mol | 0.031432 |
| $\Delta H_H$ | J/mol | 3563 |
| $\Delta S_H$ | J/K/mol | 14.865 |
| $\Delta V_H$ | J/bar/mol | -0.036069 |
| SiO$_2$ in B | molar % | 25.73 |
| Al$_2$O$_3$ in B | molar % | 1.37 |
| FeO in B | molar % | 0 (CMAS*) |
| | | 20.45 (CFMAS*) |
| MgO in B | molar % | 45.08 (CMAS*) |
| | | 37.03 (CFMAS*) |
| CaO in B | molar % | 27.82 (CMAS*) |
| | | 15.42 (CFMAS*) |
| SiO$_2$ in L | molar % | 41.65 |
| Al$_2$O$_3$ in L | molar % | 2.99 |
| FeO in L | molar % | 0 (CMAS*) |
| | | 2.03 (CFMAS*) |
| MgO in L | molar % | 54.99 (CMAS*) |
| | | 52.93 (CFMAS*) |
| CaO in L | molar % | 0.36 (CMAS*) |
| | | 0.40 (CFMAS*) |
| $X_B$ in pyrolite | | 0.1680 |

* C(F)MAS corresponds to CaO-(FeO)-MgO-Al$_2$O$_3$-SiO$_2$ system

**Extended Data Table 2.** Conditions and results of numerical experiments.

| Model | Mantle temp.[a] (K) | Mantle press.[a] (GPa) | Ambient water (wt%) | Anomaly water (wt%) | Anomaly temp. (K) | Anomaly size[b] (km) | $\frac{X_B}{X_B+X_L}$ | Results, Figures |
|---|---|---|---|---|---|---|---|---|
| diapir59 | 1773 | 12 | 0.0348 | 0.523 | -100 | 20 | 0.168 | partially molten plume with melt lens on top, Fig. 2 |
| diapir60 | 1923 | 22 | 0.0348 | 0.175 | +200 | 20 | 0.118 | partially molten plume below the MTZ producing small melt diapir inside the MTZ, Fig. 4 |
| diapir61 | 1923 | 23 | 0.0348 | 0.175 | +200 | 20 | 0.118 | partially molten plume, small melt diapir |
| diapir62 | 1773 | 12 | 0.0348 | 0.384 | 0 | 50[c] | 0.168 | MTZ overturn, large partially molten plume with melt lens on top, Extended Data Fig. 3 |
| diapir63 | 1773 | 12 | 0.351 | 0.701 | 0 | 50[c] | 0.168 | MTZ overturn, large partially molten plume with porosity waves inside |
| diapir64 | 1773 | 12 | 0.349 | 0.839 | -100 | 20 | 0.168 | MTZ overturn, porosity waves ahead of large partially molten plume |
| diapir65[e] | 1873 | 9.5 | 0.0349 | 0.387 | -500 | 60×20[d] | 0.150[g] | solid and then partially molten hydrous plumes with melt lens on top, small melt diapirs |
| diapir67[f] | 1873 | 9.5 | 0.0349 | 0.387 | -500 | 60×20[d] | 0.150[g] | solid and then partially molten hydrous plumes with melt lens on top, small melt diapirs, Extended Data Fig. 5 |
| diapir69 | 1773 | 21.5 | 0.0348 | 0.384 | -100 | 20 | 0.218 | stagnant solid drip core above the lower MTZ boundary, hydrous solid plume inside the MTZ |
| diapir70[f] | 1873 | 9.5 | 0.0349 | 0.386 | -500 | 60×20[d] | 0.150[g] | unfinished, solid state hydrous plumes inside the MTZ |
| diapir71 | 1873 | 21.5 | 0.0348 | 0.382 | -100 | 20 | 0.268 | stagnant drip core with partial melt above the lower MTZ boundary, hydrous solid plume inside the MTZ |
| diapir72 | 1773 | 12 | 0.227 | 0.716 | -100 | 20 | 0.168 | porosity waves ahead of partially molten plume |
| diapir73 | 1873 | 21.5 | 0.0348 | 0.382 | -500 | 20 | 0.268 | stagnant drip core with partial melt above the lower MTZ boundary, hydrous solid plume inside the MTZ |
| diapir74[f] | 1773 | 9.5 | 0.0349 | 0.392 | -100 | 60×20[d] | 0.150[g] | solid and then partially molten heterogeneous hydrous plume with melt lens on top, Extended Data Fig. 6 |
| diapir75 | 1773 | 12 | 0.220 | 0.709 | -100 | 20 | 0.168 | porosity waves ahead of partially molten plume, Extended Data Fig. 4 |
| diapir76 | 1773 | 12 | 0.217 | 0.705 | -100 | 20 | 0.168 | porosity waves ahead and inside of partially molten plume |
| diapir77 | 1873 | 21.5 | 0.0348 | 0.381 | -100 | 40 | 0.336 | sinking drip core, melt lens at the lower MTZ boundary, small melt diapirs, Fig. 4 |
| diapir78 | 1923 | 21.5 | 0.0348 | 0.384 | -100 | 20 | 0.218 | stagnant solid drip core above the lower MTZ boundary, hydrous solid plume inside the MTZ |
| diapir80 | 1923 | 22 | 0.0348 | 0.175 | +200 | 40 | 0.118 | partially molten plume below the MTZ with porosity waves inside evolving into solid plume inside the MTZ with melt lens on top, Extended Data Fig. 7 |
| diapir82 | 1773 | 12 | 0.0348 | 0.525 | -100 | 40 | 0.118 | partially molten plume with melt lens on top |
| diapir83[e] | 1773 | 9.5 | 0.0174 | 0.371 | -300 | 80×40[h] | 0.118 | solid and then partially molten heterogeneous hydrous plume |
| diapir84[f] | 1773 | 9.5 | 0.0174 | 0.372 | -300 | 80×40[h] | 0.118 | solid and then partially molten heterogeneous hydrous plume |
| diapir85[e] | 1773 | 9.5 | 0.0174 | 0.370 | -300 | 80×40[h] | 0.168 | solid and then partially molten heterogeneous hydrous plume |
| diapir86[f] | 1773 | 9.5 | 0.0174 | 0.370 | -300 | 80×40[h] | 0.168 | unfinished, solid hydrous plume |
| diapir87 | 1873 | 21.5 | 0.0348 | 0.384 | -100 | 40 | 0.218 | stagnant flattened solid drip core above the lower MTZ boundary, thin partially molten layer at the lower MTZ boundary, hydrous solid plume inside the MTZ |
| diapir89 | 1923 | 22 | 0.0348 | 0.175 | +200 | 40 | 0.168 | partially molten plume below the MTZ with porosity waves inside evolving into solid plume core slowly penetrating the MTZ and small melt diapir rising inside the MTZ, Extended Data Fig. 7 |
| diapir90 | 1773 | 12 | 0.0348 | 0.210 | 0 | 50[c] | 0.168 | MTZ overturn, large solid hydrous plume |
| diapir91 | 1773 | 12 | 0.0348 | 0.122 | 0 | 50[c] | 0.168 | MTZ overturn, large solid hydrous plume |

[a] mantle temperature and pressure boundary conditions on the top of the model
[b] diameter of the sharp anomaly for $\frac{X_B}{X_B+X_L}$, diameter of Gaussian anomaly for $X_H$ and temperature is 20 km (unless indicated differently)
[c] homogeneously hydrated MTZ
[d] elliptic anomaly with 60×20 diameters for sharp $\frac{X_B}{X_B+X_L}$ anomaly as well for Gaussian anomaly for $X_H$ and temperature
[e] 100×400 km² model with resolution 1×1 km²
[f] 100×400 km² model with resolution 0.5×0.5 km²
[g] ambient composition: $\frac{X_B}{X_B+X_L} = 0.15$
[h] elliptic anomaly with 80×40 km² km size for sharp $\frac{X_B}{X_B+X_L}$ anomaly and 40×20 km² size for Gaussian anomaly for temperature and $X_H$